\documentclass[%
twocolumn,
superscriptaddress,
showpacs,
altaffilletter,
nofootinbib,
 amsmath,
 amssymb,
aps,
prd,
floatfix,
]{revtex4-2}

\usepackage{amsmath}
\usepackage{graphicx}
\usepackage{float}
\usepackage{subcaption}
\usepackage{caption}
\usepackage{bm, upgreek}
\usepackage{hyperref}
\usepackage{autonum}

\newcommand{\ub}{\bm{\upbeta}}
\newcommand{\uo}{\bm{\upomega}}
\newcommand{\B}{\mathbf{B}}
\newcommand{\E}{\mathbf{E}}
\newcommand{\Spin}{\mathbf{S}}
\newcommand{\D}[2]{\frac{\mathrm{d}#1}{\mathrm{d}#2}}
\newcommand{\DD}[2]{\frac{\mathrm{d}^2 #1}{\mathrm{d} #2^2}}

\DeclareUnicodeCharacter{2212}{-}

\begin{document}

\title{Analytical Estimations of the Chromaticity and Corrections to the Spin Precession Frequency in Weak Focusing Magnetic Storage Rings}
\author{On Kim}
\email[Corresponding author, ]{bigstaron9@ibs.re.kr}
\affiliation{Center for Axion and Precision Physics Research, Institute for Basic Science, Daejeon 34051, Republic of Korea}
\author{Yannis K. Semertzidis}
\affiliation{Center for Axion and Precision Physics Research, Institute for Basic Science, Daejeon 34051, Republic of Korea}
\affiliation{Department of Physics, Korea Advanced Institute for Science and Technology, Daejeon 34141, Republic of Korea}
%\date{\today}

\begin{abstract}
	Understanding beam and spin dynamics are fundamental in accelerator-based experiments concerning polarized beams. And there has been growing interest in having more precise estimations of the beam and spin dynamics variables, as more high precision particle physics experiments in the Intensity Frontier appear. This paper provides analytical estimations of some of the important variables such as beam transverse chromaticities and corrections to the spin precession frequency  in the simplest type of particle accelerator: a circular magnetic storage ring with weak vertical focusing. We attempt to precisely obtain the next order approximations from the small betatron oscillations or the momentum dispersion, verified by high precision spin tracking simulation. We also discuss a potential way to suppress the corrections to the spin precession frequency, which relevant experiments may find beneficial.
\end{abstract}

\maketitle

%\tableofcontents

\section{Introduction}
Relativistic beam dynamics is a fundamental concept in accelerator physics, for storing the charged beam in particle accelerators and to conduct particle physics experiments, and has been naturally well described in many studies\cite{Weng1992, Edwards2008, Conte2008, Berz2015, Lee2019}. The Hill's equations for transverse betatron motions and transfer matrix techniques are essential tools to analytically evaluate the lattice parameters, and usually they provide more than enough accuracy for the beam phase space or acceptance, since they are well controlled below the tolerance required by the experiment. However, in certain cases, such as in extremely high precision experiments in the Intensity Frontier, it becomes crucial to anatomize the beam and spin motions of a polarized beam with much higher accuracy to understand and control the systematic effects. Examples of such experiments are the muon $g-2$ experiments\cite{Bennett2006, Abi2021, Abe2019}, the storage ring proton/deuteron electric dipole moment (EDM) experiment\cite{Anastassopoulos2016, Omarov2021} (also high precision tools that were developed to control and measure the spin motions in JEDI collaboration\cite{Eversmann2015, Guidoboni2016, Hempelmann2017, Saleev2017}) or several recently proposed storage ring experimental concepts seeking for exotic dark matter signals\cite{Chang2019, Graham2021, Kim2021}.

The compact size of a storage ring with relatively simple, highly symmetric lattice elements enables us to not only store polarized small emittance beams, but to analytically estimate the beam and spin motions precisely. This paper describes the beam and spin dynamics in ideal circular magnetic storage rings with weak magnetic or electric vertical focusing. The main physical quantities we aim to evaluate are the transverse betatron chromaticities and the spin precession frequency (the so-called $g-2$ frequency). In Sec. \ref{sec:beam_dynamics}, we solve for the betatron motions and the corresponding chromaticities for both magnetic or electric focusing, verified by a high precision tracking simulation. There are two types of chromaticities, natural and artificial, where the former originates from the dipole or quadrupole fields and the latter from non-linear fields. We only deal with the natural chromaticity throughout this paper. The focusing elements are assumed to be ideal and homogeneous everywhere in the storage ring, but it is worth pointing out that extending the result to the case of discrete lattice elements should be straightforward by obtaining piecewise solutions to the transfer matrix equations. In Sec. \ref{sec:spin_dynamics}, we evaluate the spin precession frequency and the leading order corrections to it caused by the momentum dispersion and the vertical betatron oscillation. Storage rings with electric focusing have been used in the previous muon $g-2$ experiments\cite{Bailey1979, Bennett2006} as well as the current one being conducted in Fermi National Accelerator Laboratory\cite{Albahri2021_PRAB, Albahri2021_PRD}. Finally, we briefly address an idea to suppress the correction to the spin precession frequency in Sec. \ref{sec:discussion}.

The physical quantities throughout this paper are defined in accelerator coordinates: $(x, y, s)$, where $x \equiv r - r_0$ is a radial distance of a particle from the reference orbit, $r_0$ is the storage ring radius, $y$ is a vertical distance and $s = r_0 \phi$ is a longitudinal arc length. Also note that we work with SI units.

\section{Beam dynamics and chromaticity} \label{sec:beam_dynamics}
The equations of motion are given by the relativistic Lorentz force equation\cite{Jackson1999},
\begin{align}
	\D{}{t} \left( \gamma m \mathbf{v} \right) &= q \left( \E + \mathbf{v} \times \B \right). \label{eq:LorentzForce}
\end{align}
The Lorentz factor $\gamma$ can be regarded as constant in time, $\dot{\gamma} = q \ub \cdot \E /(mc) \approx 0$, with the normal paraxial approximation ($\ub \cdot \E \approx 0$) in the storage ring. Rewriting the first equation for each vector component, we obtain
\begin{align}
	\ddot{r} - r \dot{\phi}^2 &= \frac{q}{\gamma m} ( \mathbf{E} + \mathbf{v} \times \mathbf{B} )_x, \\
	\ddot{y} &= \frac{q}{\gamma m} ( \mathbf{E} + \mathbf{v} \times \mathbf{B} )_y, \\
	2\dot{r}\dot{\phi} + r \ddot{\phi} &= \frac{q}{\gamma m} ( \mathbf{E} + \mathbf{v} \times \mathbf{B} )_s, \label{eq:Lorentz_s}
\end{align}
where the subscripts $x, y$ and $s$ stand for $x$-, $y$- and $s$-component of the vector in accelerator coordinates, respectively.

Representing the first equation in terms of $x = r - r_0$ and converting the time derivatives to ones with respect to the azimuth $\phi$, the first two equations become
\begin{align}
	\DD{x}{\phi} &= \frac{q}{\gamma m \dot{\phi}^2} (\E + \mathbf{v} \times \B)_x + r, \label{eq:x_eq0} \\
	\DD{y}{\phi} &= \frac{q}{\gamma m \dot{\phi}^2} (\E + \mathbf{v} \times \B)_y. \label{eq:y_eq0}
\end{align}
Note that the above equations are exact as long as the momentum of the particle is conserved. Normally the electric and magnetic fields in accelerators are functions of azimuth $\phi$, but in this paper we shall consider the fields to be homogeneous and continuous over the storage ring. We will look into each case of magnetic and electric weak focusing, respectively.

Before proceeding, it is convenient to define some useful notations. Let $\omega_{c0}$ be the cyclotron frequency for the reference particle which is given as
\begin{align}
	\omega_{c0} = \frac{v_0}{r_0} = \frac{q B_0}{\gamma_0 m}.
\end{align}
Unless specified, the subscript `$0$' basically stands for a constant physical quantity defined for the reference particle travelling in the design orbit. The effect of momentum dispersion can be quantified by a normalized momentum offset $\delta$:
\begin{align}
	\delta \equiv \frac{\Delta p}{p_0}, \qquad \frac{\Delta \gamma}{\gamma_0} = \left( 1 - \frac{1}{\gamma_0^2} \right) \delta = \beta_0^2 \delta.
\end{align}

This paper focuses on obtaining accurate analytic expressions for the transverse tunes and natural chromaticities in azimuthally symmetric ideal magnetic storage rings, with minimal approximations and a formalism that is easily applicable to higher order effects. Readers who are interested in dynamics in an electric storage rings or particular non-linear effects are directed to Refs.~\cite{Mane2008, Mane2012}.

	\subsection{Magnetic Focusing}
The magnetic field can be represented in dipole and weak focusing quadrupole terms,
\begin{align}
	\B = B_0 \hat{y} - n\frac{B_0}{r_0} (y \hat{x} + x \hat{y}),
\end{align}
where $0 < n < 1$ is the weak focusing field index. Although this ideal quadrupole field does not satisfy Maxwell's equations in cylindrical coordinates, it approximately describes the field well enough, especially when the storage ring radius is sufficiently larger than the transverse beam acceptances: $(x, y)/r_0 \ll 1$. Also, sometimes higher order multipoles are artificially suppressed by geometrical properties of the focusing elements (for instance, see Table~5 of Ref.~\cite{Semertzidis2003} for the electric quadrupole focusing).

		\subsubsection{Horizontal Chromaticity}
Equation~\eqref{eq:x_eq0} becomes
\begin{align}
	\DD{x}{\phi} &= \frac{q}{\gamma m \dot{\phi}^2} (-v_s B_y) + r \\
	&= -\frac{q B_0}{\gamma m} \frac{r}{\dot{\phi}} \left( 1 - n\frac{x}{r_0} \right) + r, \label{eq:dd_x_phi_mf}
\end{align}
using $v_s = r \dot{\phi}$.
In the simplest case without the momentum offset $\delta$, the above expression reduces to
\begin{align}
	\DD{x}{\phi} &\approx -(1-n) x
\end{align}
up to the first order of $x$, leading to the reference horizontal tune $\nu_{x0} = \sqrt{1 - n}$.
To jump into the next precision level, however, one should notice the complexity that originates from the term $\dot{\phi}$. If the horizontal position $x$ is constant in time, so is the angular frequency, namely,
\begin{align}
	\dot{\phi} (x = x_e) &= \frac{v_s}{r_0 + x_e} = \frac{v_0 \left( 1 + \frac{\delta}{\gamma_0^2} \right)}{r_0 \left( 1 + \frac{x_e}{r_0} \right)} \\
	&\approx \omega_{c0} \left( 1 + \frac{\delta}{\gamma_0^2} \right) \left( 1 - \frac{x_e}{r_0} \right), \label{eq:phidot_at_xe}
\end{align}
where $x_e$ is an equilibrium radial position which should be determined by $\delta$.
A simple force equilibrium model gives the relation between the momentum and the magnetic field, $p = q B r$, and it immediately follows that
\begin{align}
	\frac{p}{p_0} = \left( 1 - n \frac{x_e}{r_0} \right) \left( 1 + \frac{x_e}{r_0} \right).
\end{align}
Therefore, one finds
\begin{align} \label{eq:x_e}
	x_e \approx r_0 \frac{\delta}{1-n}
\end{align}
to the first order of $\delta$.
Now let us find a general expression for $\dot{\phi}$ in terms of $x$ and $\delta$. Multiplying both sides of Eq. \eqref{eq:Lorentz_s} by $r$, one obtains
\begin{align}
	\D{}{t} \left( r^2 \dot{\phi} \right) &= \frac{q}{\gamma m} r ( v_x B_y - v_y B_x ) \\
	&= \frac{q B_0}{\gamma m} \left( r \dot{r} - n \frac{r}{r_0} x \dot{x} + n \frac{r}{r_0} y \dot{y} \right) \\
	&\approx \frac{q B_0}{\gamma m} \D{}{t} \left( \frac{r^2}{2} - \frac{n x^2}{2} - \frac{n x^3}{3 r_0} \right),
\end{align}
where the term related to the vertical position $y$ is neglected in the last expression because it does not affect the horizontal betatron motion unless the coupled term survives in average, which is precisely the tune resonance that we always assume to be avoided.

Hence, the general expression for $\dot{\phi}$ looks like
\begin{align} \label{eq:phidot_general}
	\frac{\dot{\phi}}{\omega_{c0}} &\approx \left( 1 - \beta_0^2 \delta \right) \left( \frac{1}{2} - \frac{n x^2}{2 r^2} - \frac{n x^3}{3 r_0 r^2} + \frac{C(\delta)}{r^2} \right),
\end{align}
where $C(\delta) \approx C_0 + C_1 \delta$ is a constant which only depends on $\delta$. The above expression must coincide with Eq. \eqref{eq:phidot_at_xe} where $x_e$ is substituted by Eq. \eqref{eq:x_e}. That determines the two coefficients as $C_0 = r_0^2/2$ and $C_1 = r_0^2$. Finally, plugging this result into Eq. \eqref{eq:dd_x_phi_mf} and substituting $x$ with $\eta \equiv x - x_e$ which is the horizontal betatron oscillation with respect to the equilibrium chromatic orbit, we arrive to
\begin{align} \label{eq:eta_MF}
	\DD{\eta}{\phi} + \left( 1-n - \frac{n (1+n)}{1-n} \delta \right) \eta = 0.
\end{align}
The off-momentum horizontal tune $\nu_x$ is given as
\begin{align}
	\nu_x \approx \nu_{x0} \left( 1 - \frac{n (1+n)}{2(1-n)^2} \delta \right),
\end{align}
from which the horizontal chromaticity $\xi_x$ can be derived.
\begin{align}
	\xi_x \equiv \frac{\Delta \nu_x}{\delta} \approx - \nu_{x0} \frac{n (1+n)}{2(1-n)^2}.
\end{align}

		\subsubsection{Vertical Chromaticity}
The vertical chromaticity is much more straightforward to evaluate than the horizontal chromaticity. Starting from Eq. \eqref{eq:y_eq0}, it gives
\begin{align}
	\DD{y}{\phi} &= \frac{q}{\gamma m \dot{\phi}^2} (v_s B_x) \\
	&= \frac{q B_0}{\gamma m} \frac{r}{\dot{\phi}} \left( -n \frac{y}{r_0} \right). \label{eq:dd_y_phi_mf}
\end{align}
Without the momentum dispersion, it simply reduces to
\begin{align}
	\DD{y}{\phi} &\approx -n y,
\end{align}
giving the well-known expression for the reference vertical tune: $\nu_{y0} = \sqrt{n}$. Thanks to the fact that the equilibrium vertical position $y_e$ is always 0 regardless of $\delta$, one does not have to be bothered to compute Eq. \eqref{eq:phidot_general} with $y$ again. A momentum offset only shifts the horizontal equilibrium chromatic orbit, so one needs to put Eq. \eqref{eq:phidot_at_xe} into Eq. \eqref{eq:dd_y_phi_mf}, to obtain
\begin{align}
	\DD{y}{\phi} + \left( n + \frac{n (1+n)}{1-n} \delta \right) y = 0.
\end{align}
Notice that the term with $\delta$ is the same as the horizontal case with flipped sign. This leads us to the off-momentum vertical tune
\begin{align}
	\nu_y \approx \nu_{y0} \left( 1 + \frac{1+n}{2(1-n)} \delta \right),
\end{align}
where the vertical chromaticity $\xi_y$ follows as
\begin{align}
	\xi_y \equiv \frac{\Delta \nu_y}{\delta} \approx \nu_{y0} \frac{1+n}{2(1-n)}.
\end{align}

	\subsection{Electric Focusing}
The magnetic dipole and the focusing electric quadrupole fields are represented as follows.
\begin{align}
	\B = B_0 \hat{y},\qquad \E = \kappa ( x \hat{x} - y \hat{y} ),
\end{align}
where $\kappa$ is an electric field gradient. We define the field index $n$ as $\kappa r_0 / v_0 B_0$, to be consistent with the magnetic focusing case.

The procedure goes precisely the same way as in the magnetic focusing case, which does not necessarily mean that it would end up giving the same results. Nonetheless, we would like to point out that the reference tunes $\nu_{x0}, \nu_{y0}$ and the equilibrium position $x_e$ remain the same as in the magnetic focusing case, which is so straightforward that it is not derived in this paper.

		\subsubsection{Horizontal Chromaticity}
One encounters Eq. \eqref{eq:x_eq0} which becomes
\begin{align}
	\DD{x}{\phi} &= \frac{q B_0}{\gamma m \dot{\phi}^2} \left( v_0 n \frac{x}{r_0} - v_s \right) + r \\
	&= -\frac{q B_0}{\gamma m} \frac{r}{\dot{\phi}} \left( 1 - n \frac{v_0}{v_s} \frac{x}{r_0} \right) + r, \label{eq:dd_x_phi_ef}
\end{align}
The general expression for $\dot{\phi}$ should be revisited, because the field terms are different now in Eq. \eqref{eq:Lorentz_s}. In fact, it becomes simpler:
\begin{align}
	\frac{\dot{\phi}}{\omega_{c0}} &\approx \left( 1 - \beta_0^2 \delta \right) \left( \frac{1}{2} + \frac{C_0 + C_1 \delta}{r^2} \right).
\end{align}
Basically, the second and third order terms of $x$ are absent in this case. Nonetheless, it does not change the values for the coefficients $C_0$ and $C_1$, which remain $r_0^2/2$ and $r_0^2$, respectively. Plugging this into Eq. \eqref{eq:dd_x_phi_ef} and substituting $x$ with $\eta = x-x_e$, we obtain
\begin{align}
	\DD{\eta}{\phi} + \left[ 1-n - n \left( \frac{2+n}{1-n} - \frac{1}{\gamma_0^2} \right) \delta \right] \eta = 0.
\end{align}
Therefore, the horizontal chromaticity for the electric focusing is given as
\begin{align}
	\xi_x \approx - \nu_{x0} \frac{n}{2(1-n)} \left( \frac{2+n}{1-n} - \frac{1}{\gamma_0^2} \right).
\end{align}
Unlike the magnetic focusing case, the chromaticity now depends on the momentum.

\begin{table*}[t]
	\centering
	\caption{Analytical approximations of the transverse chromaticities in the magnetic and electric weak focusing storage rings.}
	\label{tab:chromaticities}
	\begin{tabular}{lcc}
		\hline\hline
		Chromaticity & Magnetic Focusing & Electric Focusing \\
		\hline
		\rule[0ex]{0pt}{5ex}Horizontal ($\xi_x$) & $- \nu_{x0} \dfrac{n (1+n)}{2(1-n)^2}$ & $- \nu_{x0} \dfrac{n}{2(1-n)} \left( \dfrac{2+n}{1-n} - \dfrac{1}{\gamma_0^2} \right)$ \\
		\rule[0ex]{0pt}{5ex}Vertical ($\xi_y$) & $\nu_{y0} \dfrac{1+n}{2(1-n)}$ & $\nu_{y0} \dfrac{1}{2} \left( \dfrac{1+n}{1-n} - \dfrac{1}{\gamma_0^2} \right)$ \\[3ex]
		\hline\hline
	\end{tabular}
\end{table*}

\begin{figure*}[t]
	\centering
	\begin{subfigure}[t]{0.46\textwidth}
		\includegraphics[width=\textwidth]{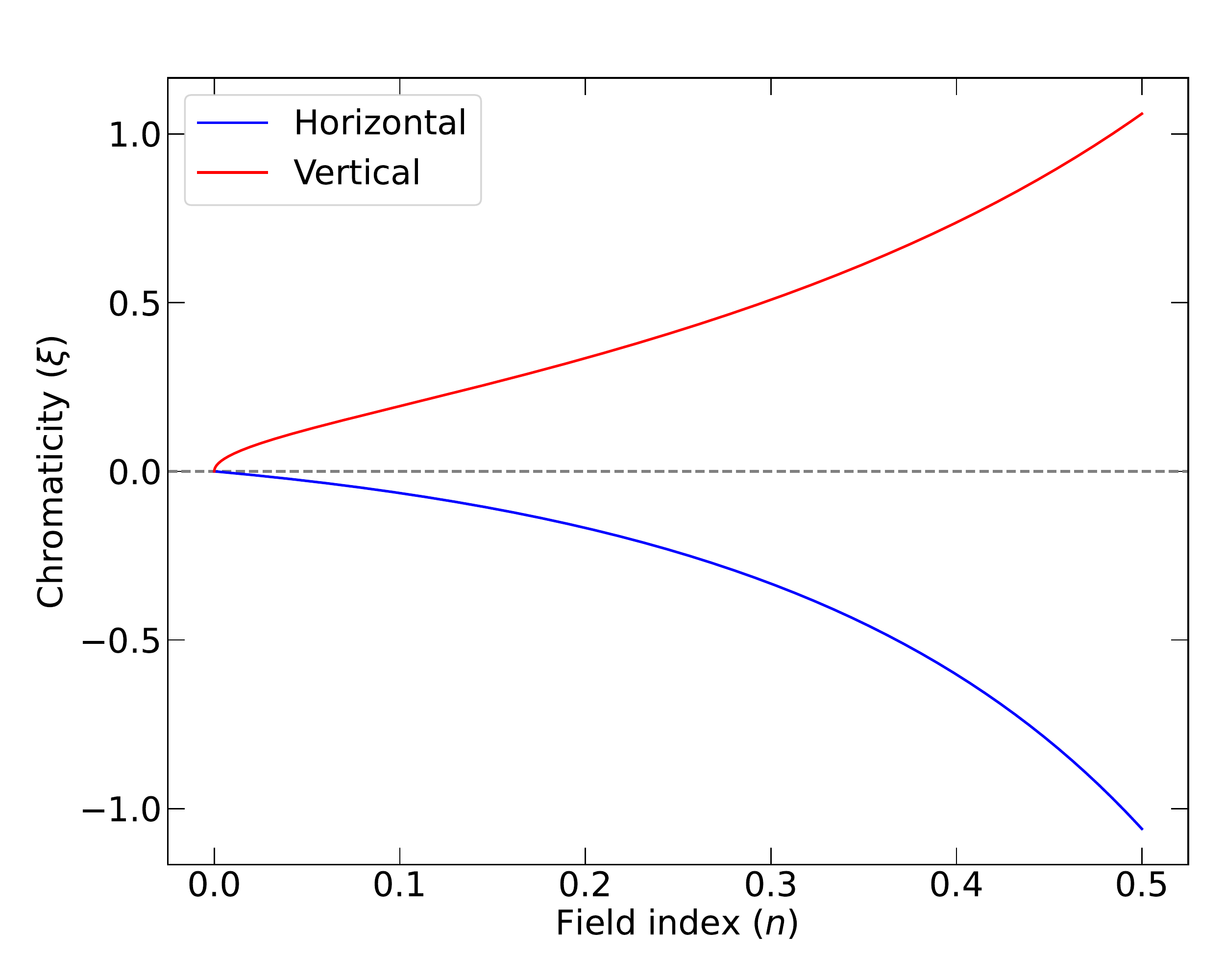}
		\caption{Magnetic focusing case.}
	\end{subfigure}
	~
	\begin{subfigure}[t]{0.46\textwidth}
		\includegraphics[width=\textwidth]{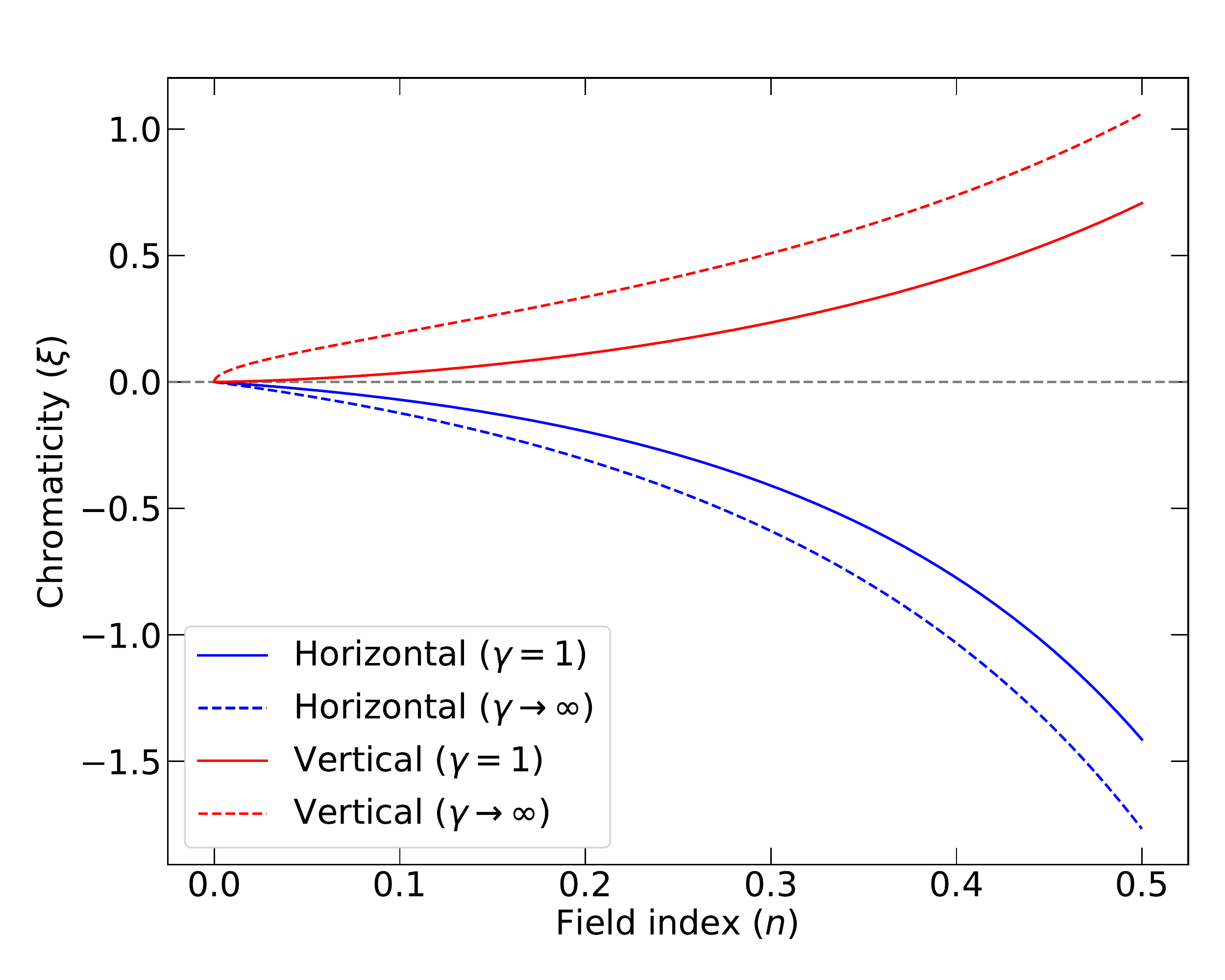}
		\caption{Electric focusing case. The solid lines indicate the low momentum limit ($\gamma = 1$) and the dashed lines indicate the high momentum limit ($\gamma \rightarrow \infty$).}
	\end{subfigure}
	\caption{Analytic approximations of the transverse chromaticities as a function of field index in weak magnetic or electric focusing storage rings. In contrast to the magnetic focusing ring, the chromaticities depend on the momentum in the electric focusing ring.}
	\label{fig:chromaticities}
\end{figure*}

		\subsubsection{Vertical Chromaticity}
The vertical equation of motion is given as
\begin{align}
	\DD{y}{\phi} &= \frac{q B_0}{\gamma m} \frac{r_0 \omega_{c0}}{\dot{\phi}^2} \left(-n \frac{y}{r_0} \right). \label{eq:dd_y_phi_ef}
\end{align}
Plugging Eq. \eqref{eq:phidot_at_xe} into the above equation, one obtains
\begin{align}
	\DD{y}{\phi} + \left( n + n \left( \frac{1+n}{1-n} - \frac{1}{\gamma_0^2} \right) \delta \right) y = 0.
\end{align}
The vertical chromaticity is therefore given as follows.
\begin{align}
	\xi_y \approx \nu_{y0} \frac{1}{2} \left( \frac{1+n}{1-n} - \frac{1}{\gamma_0^2} \right).
\end{align}

	\subsection{Verification}
We have obtained the analytical expressions for the transverse chromaticities in each case of the magnetic and electric weak focusing storage rings. The results are summarized in Table~\ref{tab:chromaticities} and their patterns as a function of the field index $n$ are shown in Fig.~\ref{fig:chromaticities}.

A high precision particle tracking simulation was conducted to verify the analytical estimations. Utilizing the 8th-order Runge-Kutta-Fehlberg algorithm\cite{Runge1895, Kutta1901, Fehlberg1969}, we conducted the numerical integration of the equations of motion and spin, without any approximation to the equations.

\begin{figure*}[t]
	\centering
	\begin{subfigure}[t]{0.46\textwidth}
		\includegraphics[width=\textwidth]{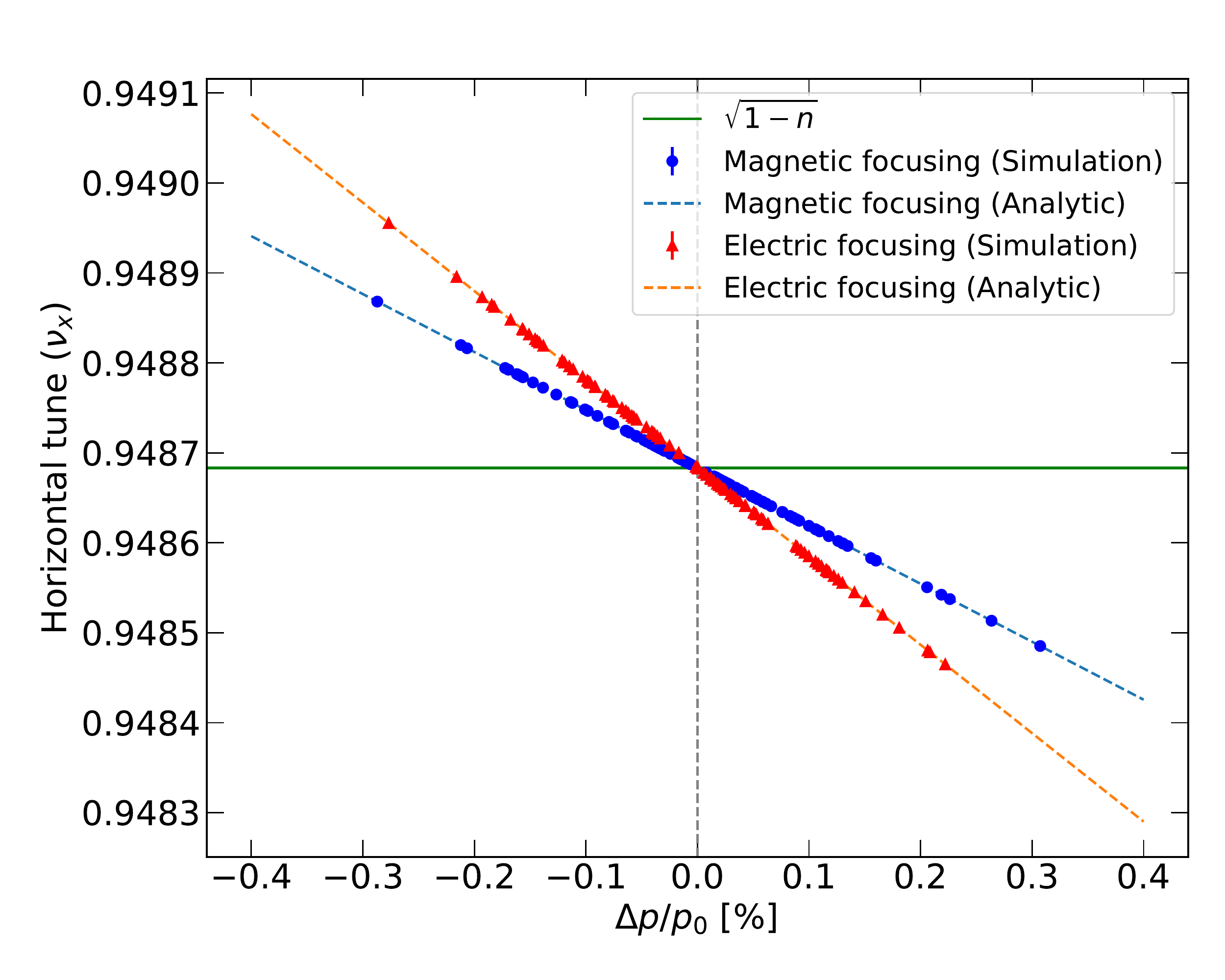}
		\caption{Horizontal tune ($\nu_x$) as a function of the fractional momentum offset.}
	\end{subfigure}
	~
	\begin{subfigure}[t]{0.46\textwidth}
		\includegraphics[width=\textwidth]{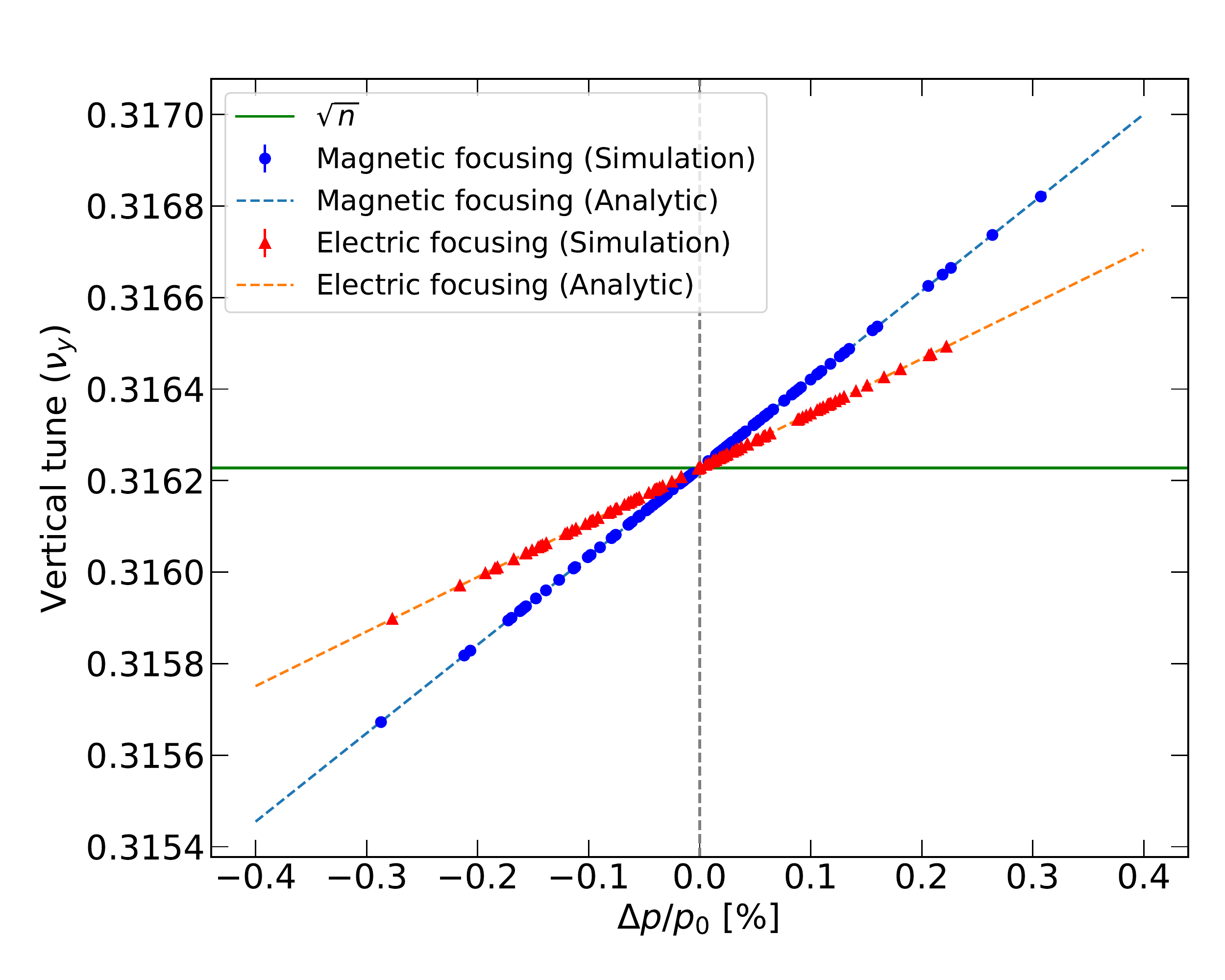}
		\caption{Vertical tune ($\nu_y$) as a function of the fractional momentum offset.}
	\end{subfigure}
	\caption{The transverse betatron tunes from the high precision tracking simulation and the analytic estimation. Each point is a betatron tune obtained by fitting the betatron oscillations from simulation, which agree well with the analytical estimations, indicated as dashed lines. The reference tunes $(\nu_{x0} = \sqrt{1-n}, \nu_{y0} = \sqrt{n})$ are shown as green horizontal solid lines in each figure. Details of the simulation setup and parameters are described in the text.}
	\label{fig:tunes_simulation}
\end{figure*}

Hundreds of protons with a momentum of around 1 GeV/c were simulated in a lattice of 10 m radius and field index $n=0.1$. The initial spatial distribution and momentum dispersion were assumed to be Gaussian around the reference orbit and momentum, with a spread of roughly $\sigma(x/r_0) \sim \sigma(y/r_0) \sim 0.2\%$ for the transverse positions and $\sigma(\Delta p/p_0) \sim 0.1\%$ for the momentum dispersion, respectively. Subsequently, the betatron oscillation for each particle was fitted in time domain to obtain the precise betatron tunes for comparison with the analytic model, which is shown in Fig.~\ref{fig:tunes_simulation}. Each point represents the horizontal or vertical betatron tunes for each particle, and the dashed lines come from the analytic estimations of corresponding chromaticities. They are consistent in both cases with the magnetic and electric focusing, verifying our high-accuracy analytic approximations.

\begin{figure*}[t]
	\centering
	\begin{subfigure}[t]{0.46\textwidth}
		\includegraphics[width=\textwidth]{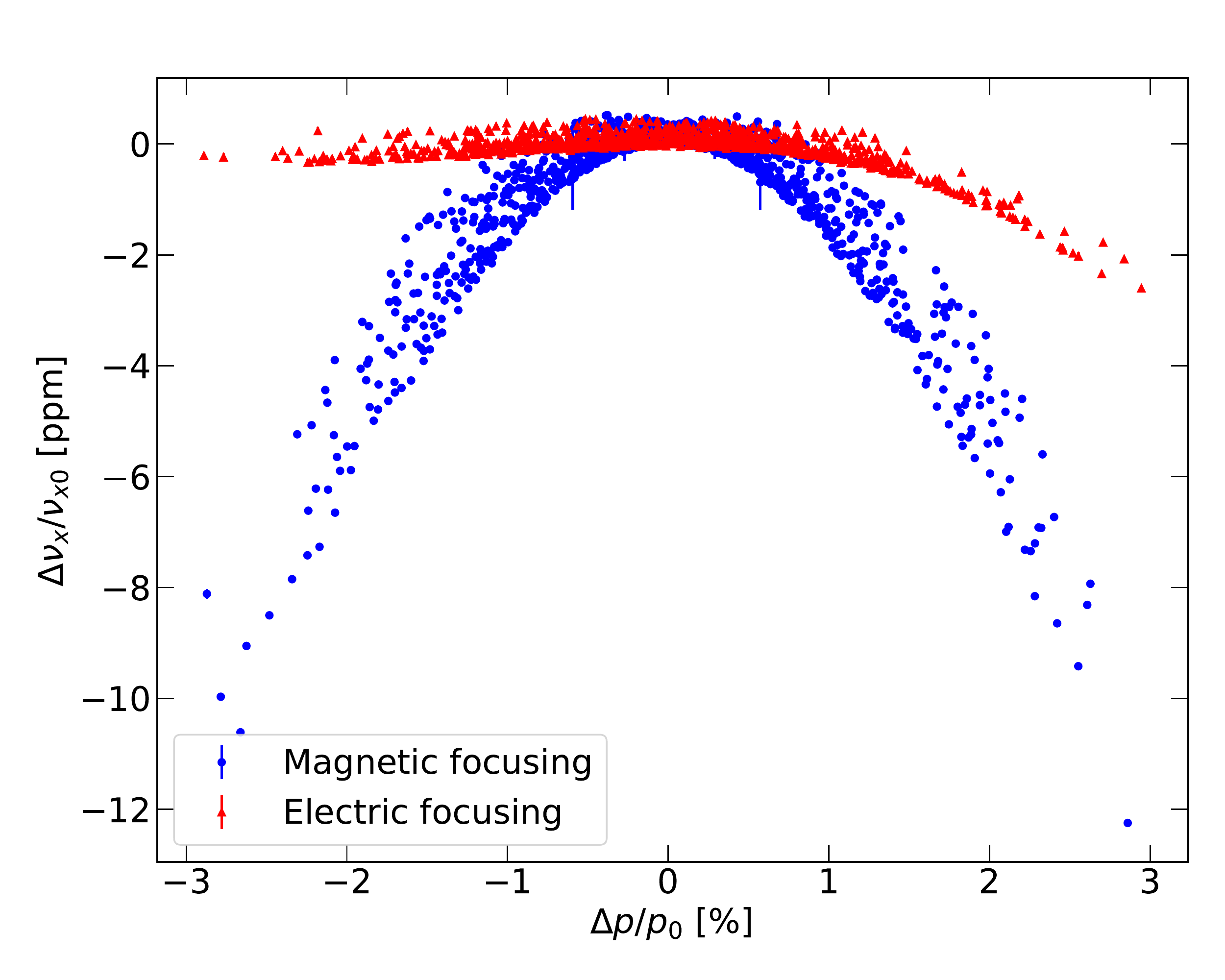}
		\caption{Relative difference of the horizontal tune ($\Delta \nu_x / \nu_{x0}$) as a function of the fractional momentum offset.}
	\end{subfigure}
	~
	\begin{subfigure}[t]{0.46\textwidth}
		\includegraphics[width=\textwidth]{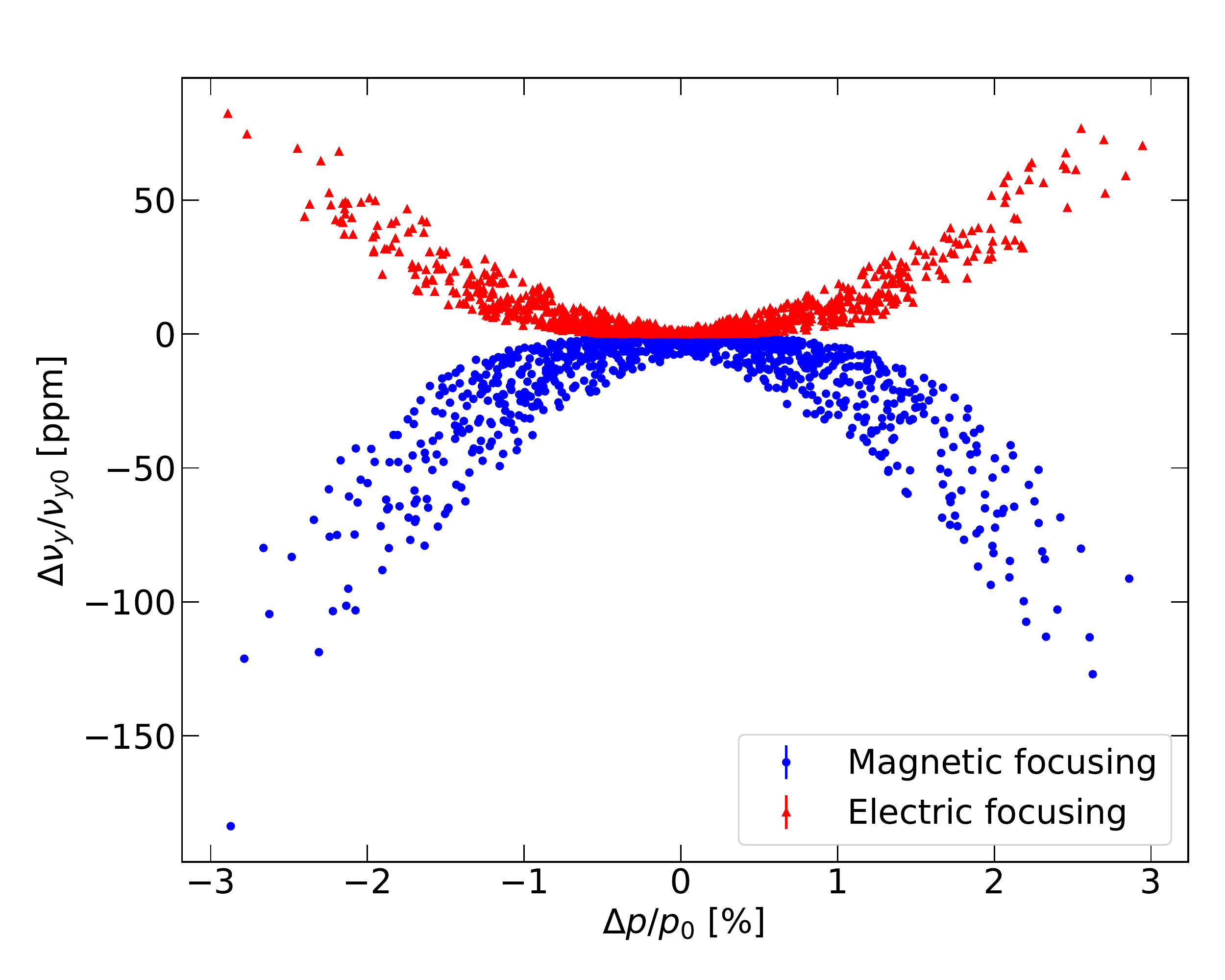}
		\caption{Relative difference of the vertical tune ($\Delta \nu_y / \nu_{y0}$) as a function of the fractional momentum offset.}
	\end{subfigure}
	\caption{The relative differences of the transverse betatron tunes between numerical and analytic estimations. The analytic estimations were obtained to the first order of the normalized momentum offset $\delta$. Details of the simulation setup and parameters are described in the text.}
	\label{fig:relative_tunes_verification}
\end{figure*}

The discrepancy of the transverse tunes, between the numerical and analytical estimations, starts to diverge in parts-per-million (ppm) level when $|\Delta p/p_0| > 1\%$, as shown in Fig.~\ref{fig:relative_tunes_verification}. The relative differences were normalized to the reference tunes: $\Delta \nu/\nu_0$. The maximum horizontal and vertical excursions were extended to 0.1~m (1\% of the storage ring radius that was used in the numerical studies), and the momentum distribution was also wider: $\sigma(\Delta p/p_0) \sim 1\%$. The parabolic patterns imply that the discrepancies are predominantly from the second-order effect of the momentum dispersion, rather than the transverse betatron amplitudes. Of course, in the presence of field errors and multipole components, the amplitude-dependent effect to the tunes can dominate over the effect of dispersion\cite{Weisskopf2019}, which is not the scope of this paper nonetheless. Typically in high precision storage ring experiments, the maximum normalized momentum offset can be controlled well below 1\%. For references, $|\Delta p/p_0|_{\max} < 0.05\%$ in the proton EDM experiment (proposed value in Ref.~\cite{Anastassopoulos2016}) and $|\Delta p/p_0|_{\max} < 0.3\%$ in the muon $g-2$ experiment\cite{Albahri2021_PRD}.

\section{Corrections to the spin precession frequency} \label{sec:spin_dynamics}
Let us turn our attention to the spin dynamics. Our goal is to obtain an accurate analytic expression for the spin precession frequency in ideal storage rings with continuous magnetic or electric quadrupole focusing. The master equation for the relativistic spin motion is given by the Thomas-Bargmann-Michel-Telegdi equation\cite{Thomas1926, Bargmann1959}:
\begin{widetext}
\begin{equation} \label{eq:omega_s}
	\D{\Spin}{t} = \uo_s \times \Spin, \qquad
	\uo_s = -\frac{q}{m} \bigg[ \left(G + \frac{1}{\gamma} \right) \B - G \frac{\gamma}{\gamma+1} (\ub \cdot \B) \ub - \left( G + \frac{1}{\gamma+1} \right) \frac{\ub \times \E}{c} \bigg],
\end{equation}
\end{widetext}
Conventionally, the magnetic anomaly $G \equiv (g-2)/2$ is often represented by $G$ for hadrons and $a$ for leptons. Note that the spin $\Spin$ is defined in the particle rest frame, whereas $\E$ and $\B$ are the electric and magnetic fields in the laboratory frame. Also, the terms related to the electric dipole moment (EDM) of a particle are neglected in the above equation. The EDM of fundamental particles or nucleons has never been detected yet, and the upper limits of the EDM for the proton\cite{Sahoo2017} and the muon\cite{Bennett2009} are, for instance, $|d_p| < 2.1 \times 10^{-25} \,e \cdot \text{cm}$ and $|d_\mu| < 1.8 \times 10^{-19} \,e \cdot \text{cm}$, respectively. Nevertheless, an interested reader is directed to an existing study on the analytical spin motion considering the EDM effect as well in circular storage rings\cite{Silenko2006, Mane2012, Fukuyama2013}.

The system of the ordinary differential equations for the spin components are given as\cite{Hoffstaetter2002}
\begin{align} \label{eq:spin_dynamics}
	\D{}{t} \begin{pmatrix}
		S_x \\ S_y \\ S_s
	\end{pmatrix}
	= \left( \uo_s - \frac{v_s r_0}{r} \bm{\upkappa} \times \hat{s} \right) \times \Spin
	\equiv \bm{\Omega}_s \times \Spin,
\end{align}
where $\bm{\upkappa}$ is the curvature vector for the reference orbit of the given lattice. In our case of a simple circular storage ring, it is explicitly given as $\bm{\upkappa} = \hat{x}/r_0$. The second term inside the parenthesis in the right hand side of Eq. \eqref{eq:spin_dynamics} comes from the time-dependence of the unit vectors of the curvilinear coordinate system that we are working in. One arrives to the following equations for each spin component.
\begin{align}
	\dot{S}_x = \Omega_{sy} S_s - \Omega_{ss} S_y, \label{eq:Sxdot} \\
	\dot{S}_y = \Omega_{ss} S_x - \Omega_{sx} S_s, \label{eq:Sydot} \\
	\dot{S}_s = \Omega_{sx} S_y - \Omega_{sy} S_x, \label{eq:Ssdot}
\end{align}
where $\bm{\Omega}_s$ components are given as
\begin{align} \label{eq:Omega_definition}
	\Omega_{sx} = \omega_{sx}, \qquad \Omega_{sy} = \omega_{sy} + \frac{v_s}{r}, \qquad \Omega_{ss} = \omega_{ss}.
\end{align}

The spin precession with respect to the particle momentum, which is conventionally referred to as $g-2$ precession, occurs perpendicular to the dipole magnetic field and therefore in-plane of the storage ring. We aim to obtain the precise $g-2$ frequency in successive subsections, by solving for an in-plane spin component, say $S_s$, for both the magnetic and electric focusing cases. The initial polarization is naturally assumed to be longitudinal. Then we will derive analytical formulae of the corrections to the spin precession frequency from two independent origins: the momentum dispersion and the pitch motion. Throughout this paper, we denote them as a ``dispersion correction'' $C_d$ and a ``pitch correction'' $C_p$, respectively.

Before we proceed, it is again worth computing the $g-2$ frequency of the reference particle. Assuming an uniform and homogeneous vertical magnetic dipole in the storage ring, it is given by
\begin{align}
	\omega_{a0} = \frac{q}{m} G B_0 = G \gamma \, \omega_{c0}.
\end{align}
Notice that we just obtained the spin tune for the reference particle: $\nu_{s0} \equiv \omega_{a0}/\omega_{c0} = G \gamma$.

Likewise in the previous section, we describe the spin motions short of the spin-orbit resonance, which can be avoided by adjusting the betatron and spin tunes away from those satisfying the following relation: $N_s \nu_s + N_x \nu_x + N_y \nu_y + N_\text{sync} \nu_\text{sync} = N$ where $N$s are integers\cite{Mane2005, Conte2008}. In particular, the low-order resonances with relatively small $N$s are necessarily avoided. Studies on the spin-orbit resonance in terms of amplitude-dependent spin tune are well established\cite{Barber2001, Hoffstaetter2002}, and they are beyond the scope of this paper.

	\subsection{Magnetic weak focusing}

		\subsubsection{Dispersion Correction}
The dominant effect of a momentum offset is, as we have seen in the previous section, that the radial equilibrium chromatic orbit shifts. It effectively changes the electric and magnetic fields that the particle experiences on average, hence affecting the spin precession rate as well. The calculation is actually quite trivial. The $g-2$ frequency for the particle at a radial equilibrium position $x_e$, with the magnetic focusing, is given as
\begin{align}
	\omega_a = \omega_{a0} \left( 1 - n \frac{x_e}{r_0} \right),
\end{align}
due to the magnetic focusing field. Using the expression for the radial equilibrium position that we obtained in Eq. \eqref{eq:x_e}, the dispersion correction $C_d$ for the magnetic focusing is given as
\begin{align}
	C_d = - \frac{n}{1-n} \delta.
\end{align}
This should be accurate up to the first order of $\delta$.

		\subsubsection{Pitch Correction}
The pitch correction is much more complicated, and it can only be accurately derived by accounting for the coupled motions of each spin component, which makes it harder to grab an intuitive picture of how it happens. It is also impossible to completely solve the coupled system of differential equations in Eq. \eqref{eq:spin_dynamics}. Nonetheless, we employ an effective technique to obtain an approximate solution of the spin precession frequency. We first obtain the second-order differential equation with respect to $S_s$, and take it as a perturbed equation from the simple harmonic oscillator differential equation. Differentiating Eq. \eqref{eq:Ssdot}, one obtains
\begin{align}
	\ddot{S}_s &= \D{}{t} \left( \Omega_{sx} S_y - \Omega_{sy} S_x \right) \\
	&= \D{}{t} \left( \Omega_{sx} S_y \right) - \dot{\Omega}_{sy} S_x - \Omega_{sy}^2 S_s + \Omega_{sy} \Omega_{ss} S_y, \label{eq:SsDdot_4terms}
\end{align}
using Eq. \eqref{eq:Sxdot}. At first glance this looks far different from the simple harmonic oscillator equation except for the third term in the right hand side. However, one can notice that the vertical spin component $S_y$ should remain fairly small during the storage, and $\Omega_{sy}^2$ is supposed to be nearly $\omega_{a0}^2$, considering $\omega_{a0}$ should take a dominant portion of $\Omega_{sy}$. In a zero pitch limit, Eq. \eqref{eq:SsDdot_4terms} is immediately restored to the harmonic oscillator equation with coefficient $\Omega_{sy}^2 = \omega_{a0}^2$. We attempt to evaluate each four terms on average to properly obtain the pitch correction.

The pitch motion is basically a vertical oscillation with respect to the reference orbit, whose first-order effect should be averaged out. The second-order terms which contain either $\sin^2(\omega_y t)$ or $\cos^2(\omega_y t)$ survive the time average and contribute to the pitch correction. Let pitch angle be $\psi = \psi_0 \cos(\omega_y t + \phi_y)$. The velocity vector can be represented as
\begin{align}
	\ub \approx \beta \psi_0 \cos(\omega_y t + \phi_y) \hat{y} + \beta \left( 1 - \frac{\psi_0^2}{2} \cos^2(\omega_y t + \phi_y) \right) \hat{s}.
\end{align}
Here the phase $\phi_y$ does not affect the final result, therefore we drop it for the rest of the section. Note that for clarity we also drop the subscript `0' indicating that the particle has the reference momentum hereafter. Integrating the vertical velocity, the vertical position becomes $y = (\psi_0 r_0/\nu_{y}) \sin(\omega_y t)$.

Now we are ready to compute each component of $\bm{\Omega}_s$. Starting from $\Omega_{sx} = \omega_{sx}$,
\begin{align}
	\omega_{sx} &= -\frac{q B_0}{m \gamma} ( G \gamma + 1 ) \left( -n \frac{y}{r_0} \right) \\
	&= \psi_0 \omega_y (G \gamma + 1) \sin(\omega_y t).
\end{align}
To proceed for the other components, we first compute $\ub \cdot \B$:
\begin{align}
	\ub \cdot \B = \psi_0 \beta B_0 \left( 1 - n \frac{x}{r_0} \right) \cos(\omega_y t).
\end{align}
Subsequently, $\omega_{sy}$ and $\omega_{ss}$ becomes
\begin{align}
	\omega_{sy} &= -\frac{q B_0}{m \gamma} \left( 1 - n \frac{x}{r_0} \right) \left[ G \gamma + 1 - \psi_0^2 G (\gamma - 1) \cos^2(\omega_y t) \right] \\
	&\approx -(\omega_{a0} + \omega_{c}) \left( 1 - n \frac{x}{r_0} \right) + \psi_0^2 \omega_{a0} \frac{\gamma-1}{\gamma} \cos^2(\omega_y t). \\
	\omega_{ss} &\approx \frac{q B_0}{m \gamma} \left( 1 - n \frac{x}{r_0} \right) \psi_0 G (\gamma - 1) \cos(\omega_y t) \\
	&\approx \psi_0 \omega_{a0} \frac{\gamma-1}{\gamma} \cos(\omega_y t).
\end{align}
We actually treated $x$ as an order of $\psi_0^2$ in the above expressions. This is purely a consequence of the pitch motion, neither the momentum dispersion nor the initial radial phase space. The slight drop in longitudinal velocity due to the pitch, $-\psi_0^2/4$ on average, leads to the slightly shifted radial equilibrium position. One can go back to Eq. \eqref{eq:dd_x_phi_mf} and reason that the equilibrium position should be given as
\begin{align} \label{eq:x_e_MF}
	x_e = -\frac{\psi_0^2}{4} \frac{r_0}{1-n}.
\end{align}

For $\Omega_{sy} = \omega_{sy} + v_s/r$, we need to compute $v_s/r$:
\begin{align} \label{eq:vs_over_r_MF}
	\frac{v_s}{r} \approx \omega_c \left( 1 - \frac{\psi_0^2}{2} \cos^2(\omega_y t) + \frac{\psi_0^2}{4} \frac{1}{1-n} \right),
\end{align}
which leads to
\begin{align} \label{eq:Omega_sy_MF}
	\Omega_{sy} \approx -\omega_{a0} \left( 1 + \frac{\psi_0^2}{4} \frac{1}{1-n} - \psi_0^2 \frac{\gamma-1}{\gamma} \cos^2(\omega_y t) \right) \\ - \omega_c \frac{\psi_0^2}{4} \cos(2\omega_y t).
\end{align}

One last quantity to evaluate Eq. \eqref{eq:SsDdot_4terms} is $S_y$, which needs to be obtained by integrating Eq. \eqref{eq:Sydot}. $S_y$ should be small, an order of $\psi_0$, considering what we have for $\Omega_{sx}$ and $\Omega_{ss}$. This again provides a technique similar to the one in Eq. \eqref{eq:Sydot} as a perturbed differential equation where the leading order solution is given when $S_s = \cos(\omega_{a0}t)$ and $S_x = -\sin(\omega_{a0}t)$. It is a simple trigonometric integration, which leads to
\begin{widetext}
\begin{align} \label{eq:S_y_MF}
	S_y \approx \psi_0 \frac{ \left[ \gamma (G \gamma + 1) \omega_y^2 - (\gamma-1) \omega_{a0}^2 \right] (\cos(\omega_y t) \cos(\omega_{a0} t) - 1) + (G \gamma^2 + 1) \omega_y \omega_{a0} \sin(\omega_y t) \sin(\omega_{a0} t) }{\gamma (\omega_y^2 - \omega_{a0}^2)}.
\end{align}
\end{widetext}
Equation~\eqref{eq:S_y_MF} looks hopelessly complicated, but what matters is to work out what term would survive in $\mathrm{d} (\Omega_{sx} S_y) / \mathrm{d}t$ on average. When averaged, we only take the average of terms related to pitch oscillation, since they contribute as the coefficient of the final harmonic-oscillator-like differential equation. Of course, this does not work when the spin precession is on resonance with the vertical betatron oscillation, for instance $\omega_{a0} \sim \omega_y$. But we shall assume the off-resonance case, as they can be easily tuned away from each other by modifying the field index.

$\Omega_{sx} S_y$ essentially contains the following three terms: $\sin(\omega_y t) \cos(\omega_y t) \cos(\omega_{a0} t)$, $\sin(\omega_y t)$ and $\sin^2(\omega_y t) \sin(\omega_{a0} t)$. Among them, the only term that survives the pitch-only-average (which we denote $\langle \cdots \rangle_y$ hereafter) after the time derivative is $\sin^2(\omega_y t)$ term. Therefore, we have
\begin{align}
	\left\langle \D{}{t} \left( \Omega_{sx} S_y \right) \right\rangle_y \approx \frac{\psi_0^2}{2} \frac{(G \gamma + 1) (G \gamma^2 + 1) \omega_y^2 \omega_{a0}^2}{\gamma (\omega_y^2 - \omega_{a0}^2)} S_s.
\end{align}
The next term in Eq.~\eqref{eq:SsDdot_4terms}, $\dot{\Omega}_{sy}$ is more straightforward.
\begin{align}
	\left\langle \dot{\Omega}_{sy} \right\rangle_y &\approx \psi_0^2 \left\langle -2 \omega_{y} \omega_{a0} \frac{\gamma-1}{\gamma} \sin(\omega_y t) + \frac{\omega_c \omega_y}{2} \sin(2\omega_y t) \right\rangle_y \\
	&= 0.
\end{align}
The remaining terms are not difficult to compute, keeping terms up to the order of $\psi_0^2$. They are given as
\begin{align}
	\left\langle \Omega_{sy}^2 \right\rangle_y &\approx \omega_{a0}^2 \left( 1 + \frac{\psi_0^2}{2} \frac{n}{1-n} - \psi_0^2 \frac{\gamma-1}{\gamma} \right),
\end{align}
and
\begin{align}
	\left\langle \Omega_{sy} \Omega_{ss} S_y \right\rangle_y &\approx -\omega_{a0}^2 \frac{\psi_0^2}{2} \frac{\gamma-1}{\gamma} \frac{\gamma (G \gamma +1) \omega_y^2 - (\gamma - 1) \omega_{a0}^2}{\gamma (\omega_y^2 - \omega_{a0}^2)} S_s.
\end{align}
Finally, plugging the results into Eq.~\eqref{eq:SsDdot_4terms}, we obtain
\begin{align}
	\ddot{S}_s \approx -\omega_{a0}^2 \left[ 1 - \frac{\psi_0^2}{2} \left( 1 - \frac{n}{1-n} + \frac{G \gamma^2 (G \gamma^2 + 2) \omega_y^2 + \omega_{a0}^2}{\gamma^2 (\omega_y^2 - \omega_{a0}^2)} \right) \right] S_s.
\end{align}
Hence, the pitch correction in the magnetic focusing storage ring becomes
\begin{align} \label{eq:Cp_MF}
	C_p = - \frac{\psi_0^2}{4} \left( 1 - \frac{n}{1-n} + \frac{G \gamma^2 (G \gamma^2 + 2) \omega_y^2 + \omega_{a0}^2}{\gamma^2 (\omega_y^2 - \omega_{a0}^2)} \right).
\end{align}

	\subsection{Electric weak focusing}

		\subsubsection{Dispersion Correction}
Despite the same radial equilibrium chromatic orbit for both magnetic and electric focusing, the effect of dispersion on the spin precession varies. It can be shown by writing down the expression for $\omega_a$ of a particle with a momentum offset $\delta$, which is given as
\begin{align}
	\omega_a &= -\frac{q}{m} \left[ G B_0 + \left( G - \frac{1}{\gamma^2 - 1} \right) \frac{\beta E_x}{c} \right] \\
	&\approx \omega_{a0} \left[ 1 - \frac{\beta^2}{G} \left( G - \frac{1}{\gamma^2 - 1} \right) n \frac{x_e}{r_0} \right],
\end{align}
which leads to the dispersion correction for the electric focusing.
\begin{align} \label{eq:C_d_EF}
	C_d \approx -\frac{\beta^2}{G} \left( G - \frac{1}{\gamma^2 - 1} \right) \frac{n}{1-n} \delta.
\end{align}
For a particle with a positive magnetic anomaly $G>0$, there exists a specific momentum which cancels this dispersion correction, the so-called \textit{magic momentum}\cite{Bailey1977, Bailey1979}:
\begin{align}
	p_m \equiv \frac{mc}{\sqrt{G}},
\end{align}
where $p_m$ is roughly 0.7 GeV/c for the proton and 3.1 GeV/c for the muon, as references. Although Eq.~\eqref{eq:C_d_EF} vanishes to the first order of $\delta$ when the particle is at the magic momentum, sometimes it is necessary to take the second order effect into account for precise measurement. We have
\begin{align}
	G - \frac{1}{\gamma^2 - 1} \approx G - \frac{1}{\gamma_0^2 - 1}  + \frac{2}{\gamma_0^2 - 1} \delta = 2 G \delta,
\end{align}
restoring our notation of the subscript `0' to indicate the reference values. Therefore, in case of the magic momentum, the second order dispersion correction becomes
\begin{align}
	C_d (p \approx p_m) \approx  -2 \beta^2 \frac{n}{1-n} \delta^2.
\end{align}
This is precisely what has been called the \textit{E-field correction} in the muon $g-2$ experiment at BNL and Fermilab\cite{Bennett2006, Albahri2021_PRAB}.

		\subsubsection{Pitch Correction}
We follow the same procedure as we did in the magnetic focusing case, but there is a subtle but substantial difference between the two cases. The total energy of the particle, although it is supposed to be conserved on average, can oscillate as a function of time, because the electric field does the work on the particle. To estimate the energy $\mathcal{E}$, we recall the Lorentz force equation again.
\begin{align}
	\D{\mathcal{E}}{t} &= q \mathbf{v} \cdot \E \\
	&\approx -\psi_0^2 q B_0 v^2 \frac{\omega_y}{\omega_c} \sin(\omega_y t) \cos(\omega_y t).
\end{align}
Integrating the above equation, one obtains
\begin{align}
	\mathcal{E} \approx \mathcal{E}_0 - \frac{\psi_0^2}{2} \frac{q B_0 v^2}{\omega_c} \sin^2(\omega_y t),
\end{align}
from which we arrive at the expression for $\delta$:
\begin{align}
	\delta = \frac{1}{\beta^2} \frac{\Delta \mathcal{E}}{\mathcal{E}_0}
	\approx - \frac{\psi_0^2}{2} \sin^2(\omega_y t).
\end{align}
Therefore, the electric focusing tends to lower the initial energy by exerting a force against the pitch motion. This affects the radial equilibrium orbit as well. In addition to Eq.~\eqref{eq:x_e_MF} which comes from the pitch motion, the effect from the shifted average dispersion should be taken into account, where $\langle \delta \rangle \approx - \psi_0^2/4$. This actually doubles the radial equilibrium position from that for magnetic focusing,
\begin{align}
	x_e^\text{E-focusing} = 2x_e^\text{B-focusing} \approx -\frac{\psi_0^2}{2} \frac{r_0}{1-n}.
\end{align}

Having this result, that $\delta$ is an order of $\psi_0^2$, let us evaluate the spin frequency components of $\bm{\Omega}_s$.
\begin{align}
	\omega_{sx} &= -\frac{q}{m} \left( G + \frac{1}{\gamma + 1} \right) \frac{\beta_s E_y}{c} \\
	&\approx \psi_0 \omega_y \left( G + \frac{1}{\gamma + 1} \right) \gamma \beta^2 \sin(\omega_y t).
\end{align}
\begin{widetext}
\begin{align}
	\omega_{sy} &= -\frac{q B_0}{m} \left[ \left( G + \frac{1}{\gamma} \right) - G \frac{\gamma-1}{\gamma} \psi_0^2 \cos^2(\omega_y t) - \left( G + \frac{1}{\gamma + 1} \right) \frac{\beta_s E_x}{c B_0} \right] \\
	&\approx -\omega_{a0} \left( 1 - \psi_0^2 \frac{\gamma-1}{\gamma} \cos^2(\omega_y t)  \right) - \omega_c \left[ 1 - \beta^2 \delta + \frac{\psi_0^2}{2} \left( G + \frac{1}{\gamma + 1} \right) \gamma \beta^2 \frac{n}{1-n} \right]
\end{align}
\end{widetext}
\begin{align}
	\omega_{ss} &= \frac{q B_0}{m} \left[ G \frac{\gamma-1}{\gamma} \psi_0 \cos(\omega_y t) - \left( G + \frac{1}{\gamma + 1} \right) \frac{\beta_y E_x}{c B_0} \right] \\
	&\approx \psi_0 \omega_{a0} \frac{\gamma-1}{\gamma} \cos(\omega_y t).
\end{align}
Also, one needs to evaluate $v_s/r$ similar to Eq.~\eqref{eq:vs_over_r_MF}:
\begin{align}
	\frac{v_s}{r} \approx \omega_c \left( 1 - \frac{\psi_0^2}{2} \cos^2(\omega_y t) + \frac{\psi_0^2}{2} \frac{1}{1-n} + \frac{\delta}{\gamma^2} \right).
\end{align}
Note that the difference from Eq.~\eqref{eq:vs_over_r_MF} comes from the pitch-induced dispersion.

Having all spin angular frequency vector components, we notice that $\Omega_{ss}$ remains the same as in the magnetic focusing, whereas $(G \gamma +1)$ has changed to $(G + 1/(\gamma + 1)) \gamma \beta^2$ for $\Omega_{sx}$, and $\Omega_{sy}$ changed in a somewhat complicated way. In our strategy for calculating the leading order $S_y$ by integrating Eq.~\eqref{eq:Sydot} with $S_s = \cos(\omega_{a0}t)$ and $S_x = -\sin(\omega_{a0}t)$, we can use a trick based on our previous calculation in Eq.~\eqref{eq:S_y_MF}; just substitute $(G \gamma +1)$ into $(G + 1/(\gamma + 1)) \gamma \beta^2$, or equivalently, $G \rightarrow G - (G+1)/\gamma^2$ in Eq.~\eqref{eq:S_y_MF}.
\begin{widetext}
\begin{align}
	S_y \approx \psi_0 \frac{ \left[ G ( \gamma^2 - 1) \omega_y^2 - (\gamma-1) \omega_{a0}^2 \right] (\cos(\omega_y t) \cos(\omega_{a0} t) - 1) + G (\gamma^2 - 1) \omega_y \omega_{a0} \sin(\omega_y t) \sin(\omega_{a0} t) }{\gamma (\omega_y^2 - \omega_{a0}^2)}.
\end{align}
\end{widetext}

The last thing to do is to evaluate the four terms in Eq.~\eqref{eq:SsDdot_4terms}. The first two terms are given as follows.
\begin{align}
	\left\langle \D{}{t} \left( \Omega_{sx} S_y \right) \right\rangle_y \approx \frac{\psi_0^2}{2} \left( G + \frac{1}{\gamma+1} \right) \frac{G (\gamma^2 - 1) \beta^2 \omega_y^2 \omega_{a0}^2}{\omega_y^2 - \omega_{a0}^2} S_s,
\end{align}
and $\left\langle \dot{\Omega}_{sy} \right\rangle_y = 0$ again holds as in the magnetic focusing case. The third and fourth terms become:
\begin{align}
	\left\langle \Omega_{sy}^2 \right\rangle_y &\approx \omega_{a0}^2 \Bigg( 1 - \psi_0^2 \frac{\gamma-1}{\gamma} + \psi_0^2 \frac{n}{1-n} \frac{\left( G + \frac{1}{\gamma + 1} \right) \gamma \beta^2 - 1}{G \gamma} \Bigg),
\end{align}
and
\begin{align}
	\left\langle \Omega_{sy} \Omega_{ss} S_y \right\rangle_y &\approx -\omega_{a0}^2 \frac{\psi_0^2}{2} \left( \frac{\gamma-1}{\gamma} \right)^2 \left( 1 + \frac{G (\gamma +1) \omega_y^2}{\omega_y^2 - \omega_{a0}^2} \right) S_s.
\end{align}
Eventually, we arrive to the pitch correction for the electric focusing.
\begin{widetext}
\begin{align} \label{eq:Cp_EF}
	C_p = -\frac{\psi_0^2}{4} \left[ 1 + \frac{G^2 (\gamma^2 - 1)^2 \omega_y^2}{\gamma^2 (\omega_y^2 - \omega_{a0}^2)} + \frac{2n - G \left\{ 1 - n + 2n (\gamma^2 - 1) \right\}}{G (1-n) \gamma^2} \right].
\end{align}
\end{widetext}
For the particle with the magic momentum that satisfies $G(\gamma^2 -1) = 1$, the pitch correction reduces to a simpler form:
\begin{align} \label{eq:Cp_EF_magic}
	C_p (p = p_m) = -\frac{\psi_0^2}{4} \left( 1 + \frac{\omega_{a0}^2}{\gamma^2 (\omega_y^2 - \omega_{a0}^2)} \right).
\end{align}

\begin{figure*}[t]
	\centering
	\begin{subfigure}[t]{0.46\textwidth}
		\includegraphics[width=\textwidth]{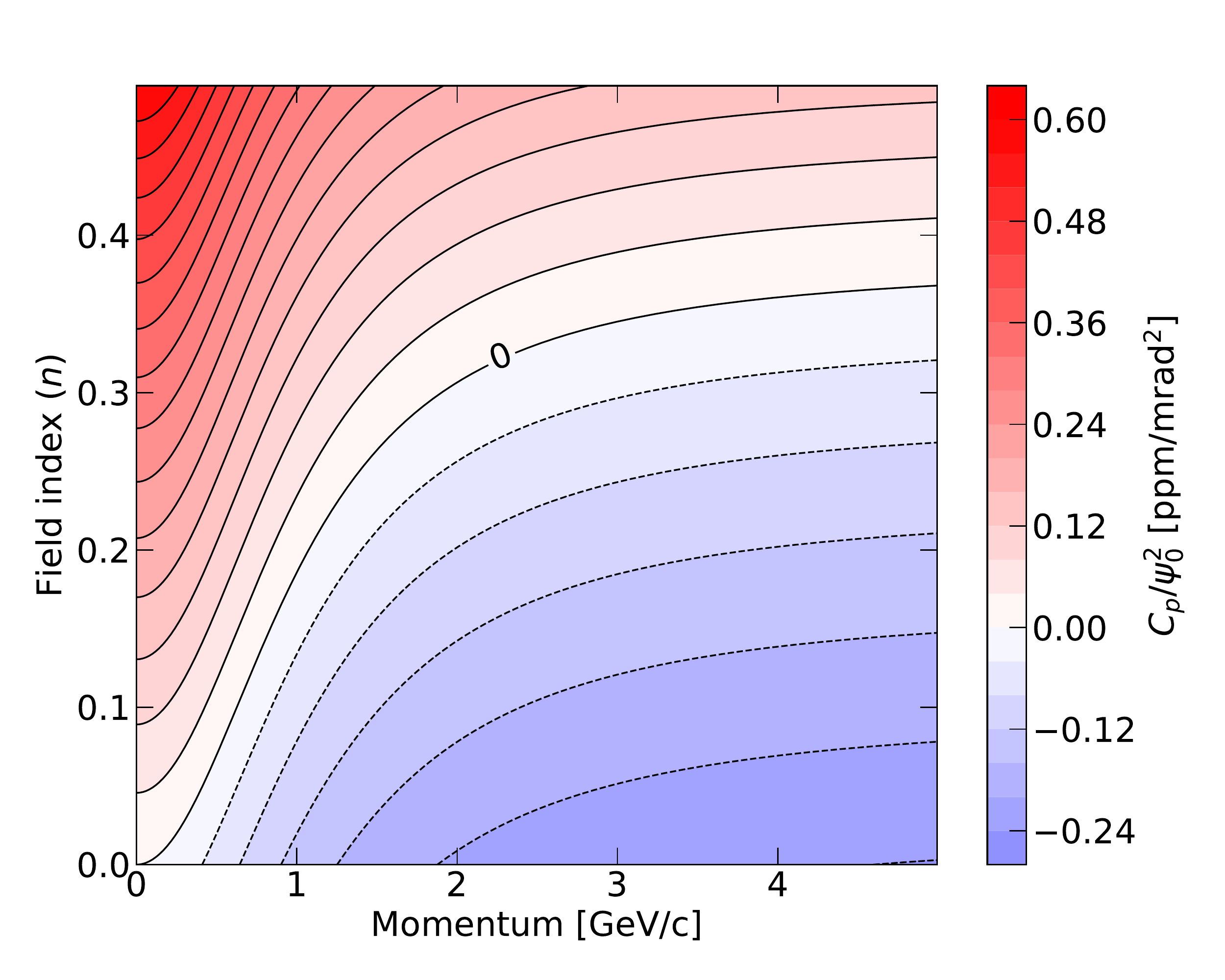}
		\caption{For protons with magnetic focusing.}
	\end{subfigure}
	~
	\begin{subfigure}[t]{0.46\textwidth}
		\includegraphics[width=\textwidth]{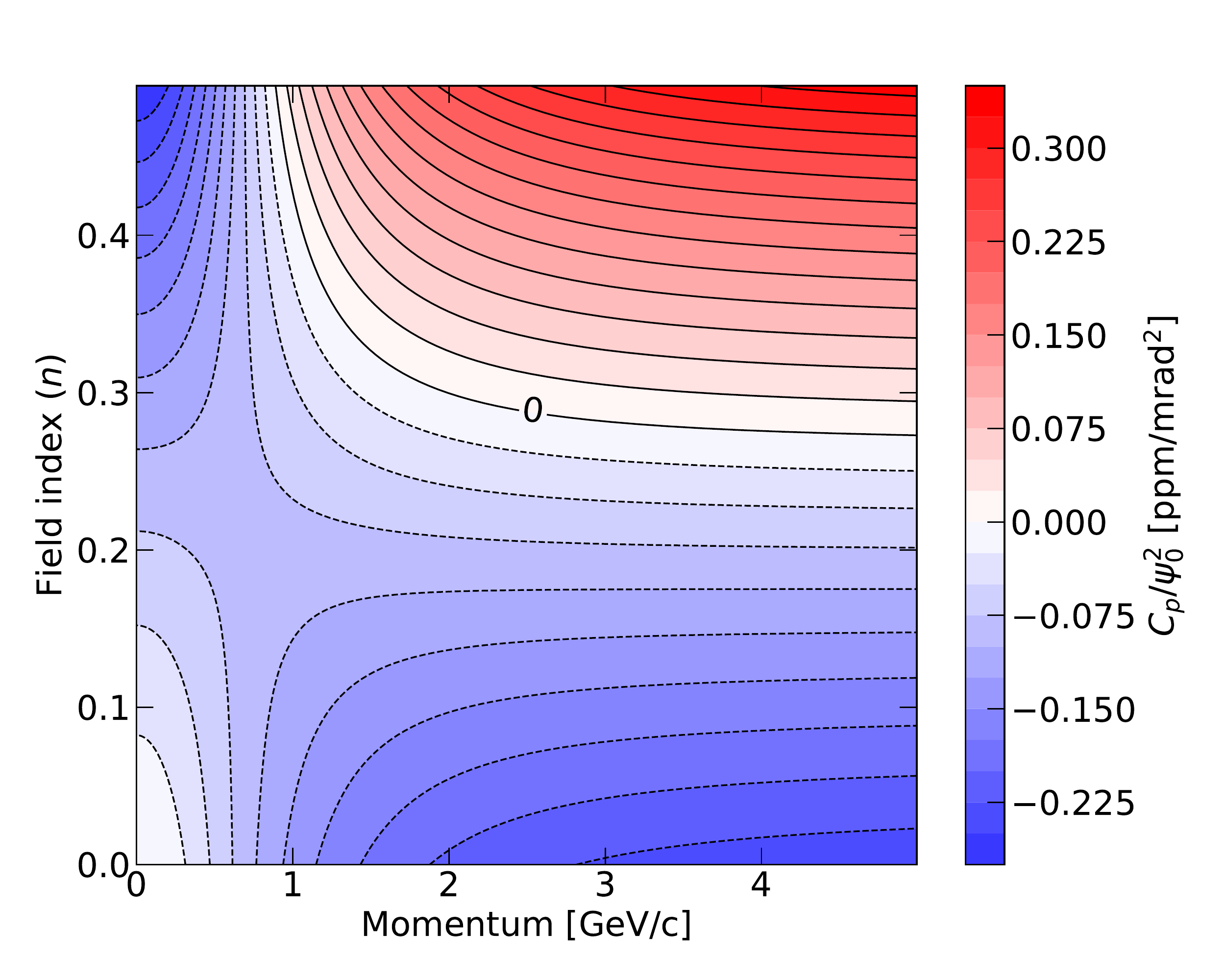}
		\caption{For protons with electric focusing.}
	\end{subfigure}
	~
	\begin{subfigure}[t]{0.46\textwidth}
		\includegraphics[width=\textwidth]{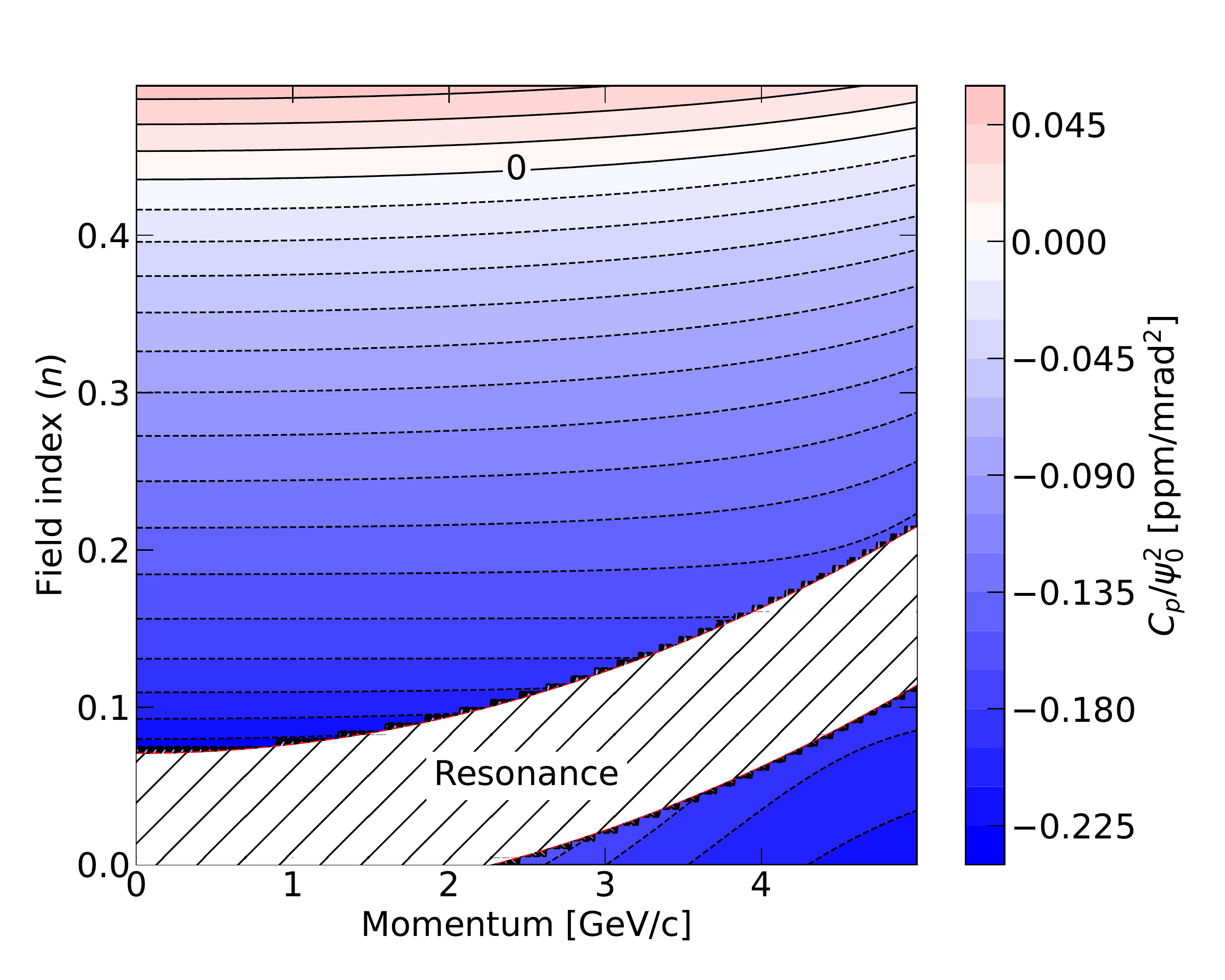}
		\caption{For deuterons with magnetic focusing.}
	\end{subfigure}
	~
	\begin{subfigure}[t]{0.46\textwidth}
		\includegraphics[width=\textwidth]{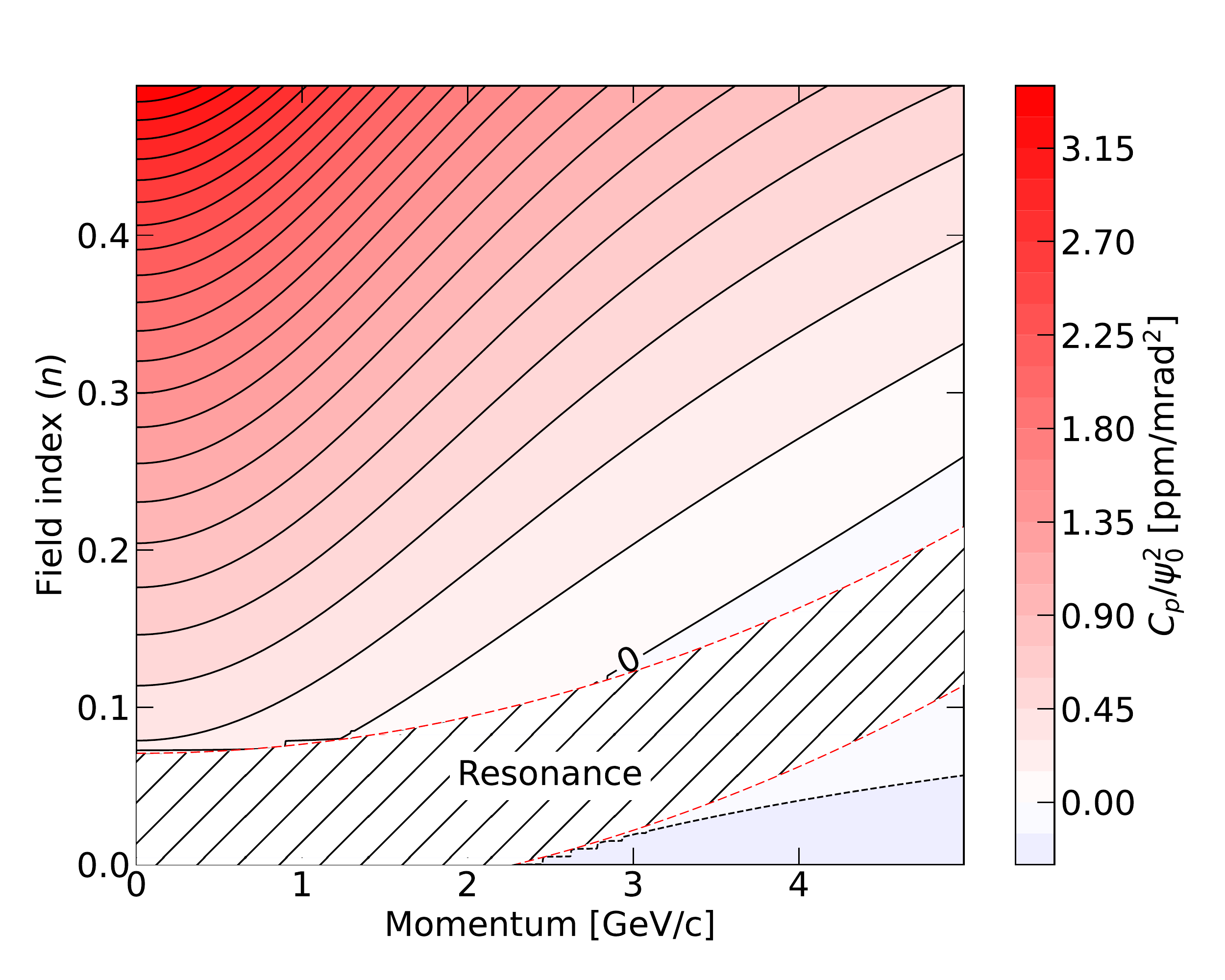}
		\caption{For deuterons with electric focusing.}
	\end{subfigure}
	\caption{The pitch correction per square pitch angle ($C_p/\psi_0^2$) in parts-per-million/$\text{mrad}^2$ as a function of the particle momentum and the weak focusing field index. Two particles, the proton with $G = 1.79$ and the deuteron with $G = -0.14$ are plotted for both magnetic and electric focusing storage rings. Specific contour lines with the label `0' indicate when the pitch correction vanishes. For deuterons, the hatched regions are near the spin resonance $\omega_a^2 = \omega_y^2$ and therefore are excluded in the plot.}
	\label{fig:Cp_analytic}
\end{figure*}

	\subsection{Verification}
\begin{table*}[t]
	\setlength{\tabcolsep}{4pt}
	\centering
	\caption{Analytical approximations of corrections to the spin precession frequency in the magnetic and electric weak focusing storage rings.}
	\label{tab:spin_corrections}
	\begin{tabular}{lcc}
		\hline\hline
		Correction & Magnetic Focusing & Electric Focusing \\
		\hline
		\rule[0ex]{0pt}{5ex}Dispersion ($C_d$) & \( \displaystyle - \dfrac{n}{1-n} \delta \) & \( \displaystyle -\frac{\beta^2}{G} \left( G - \frac{1}{\gamma^2 - 1} \right) \frac{n}{1-n} \delta \) \\
		\rule[0ex]{0pt}{5ex}Pitch ($C_p$) & \( \displaystyle -\frac{\psi_0^2}{4} \left( 1 - \frac{n}{1-n} + \frac{G \gamma^2 (G \gamma^2 + 2) \omega_y^2 + \omega_{a0}^2}{\gamma^2 (\omega_y^2 - \omega_{a0}^2)} \right) \) & \( \displaystyle -\frac{\psi_0^2}{4} \left[ 1 + \frac{G^2 (\gamma^2 - 1)^2 \omega_y^2}{\gamma^2 (\omega_y^2 - \omega_{a0}^2)} + \frac{2n - G \left\{ 1 - n + 2n (\gamma^2 - 1) \right\}}{G (1-n) \gamma^2} \right] \) \\[3ex]
		\hline\hline
	\end{tabular}
\end{table*}

Our results for magnetic and electric focusing are summarized in Table~\ref{tab:spin_corrections}. If the dispersion and pitch angle are about the same order of magnitude, then the dispersion correction $C_d$ is dominant over the pitch correction $C_p$, because the pitch correction is the second-order in terms of pitch angle. The pitch correction formulae are complicated, being the function of many parameters and the beam conditions. A common interesting question regarding the pitch correction could be under what condition it vanishes, as the dispersion correction vanishes at the magic momentum in the electric focusing. This is shown in Fig.~\ref{fig:Cp_analytic} in 2D plots as contour lines with a label `0'. The condition eliminating the pitch correction varies for different particle types, and Fig.~\ref{fig:Cp_analytic} shows it for protons and deuterons as references. Having a negative magnetic anomaly, the pitch correction for the deuteron displays quite different behavior than the proton's. Most noticeably, the spin resonance $\omega_a^2 = \omega_y^2$ appears near the low field index for the deuteron, which should necessarily be avoided. Otherwise, the pitch correction would rise uncontrollably, which is undoubtedly harming the experimental sensitivity on the spin precession frequency.

\begin{figure*}[t]
	\centering
	\begin{subfigure}[t]{0.46\textwidth}
		\includegraphics[width=\textwidth]{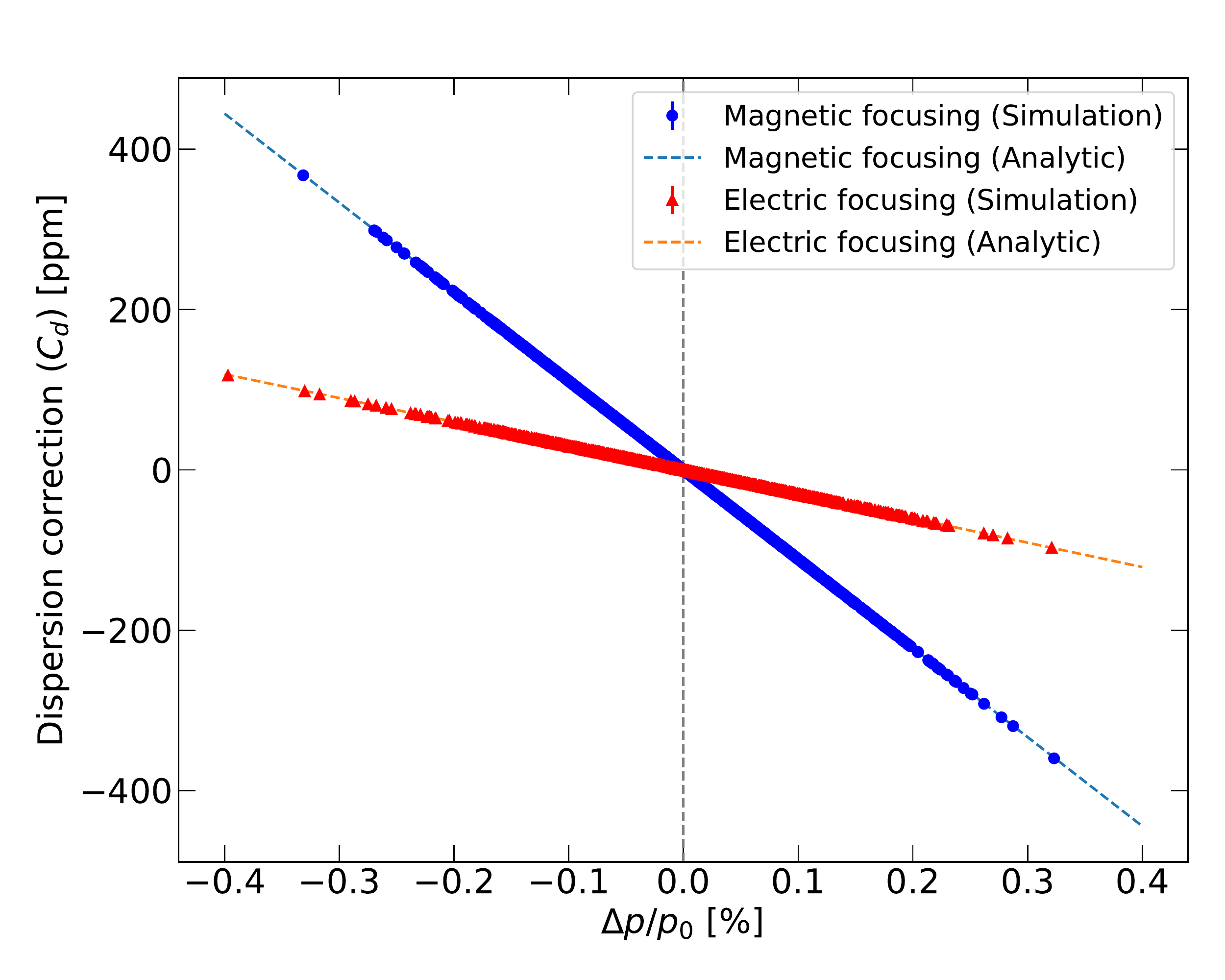}
		\caption{The dispersion correction.}
	\end{subfigure}
	~
	\begin{subfigure}[t]{0.46\textwidth}
		\includegraphics[width=\textwidth]{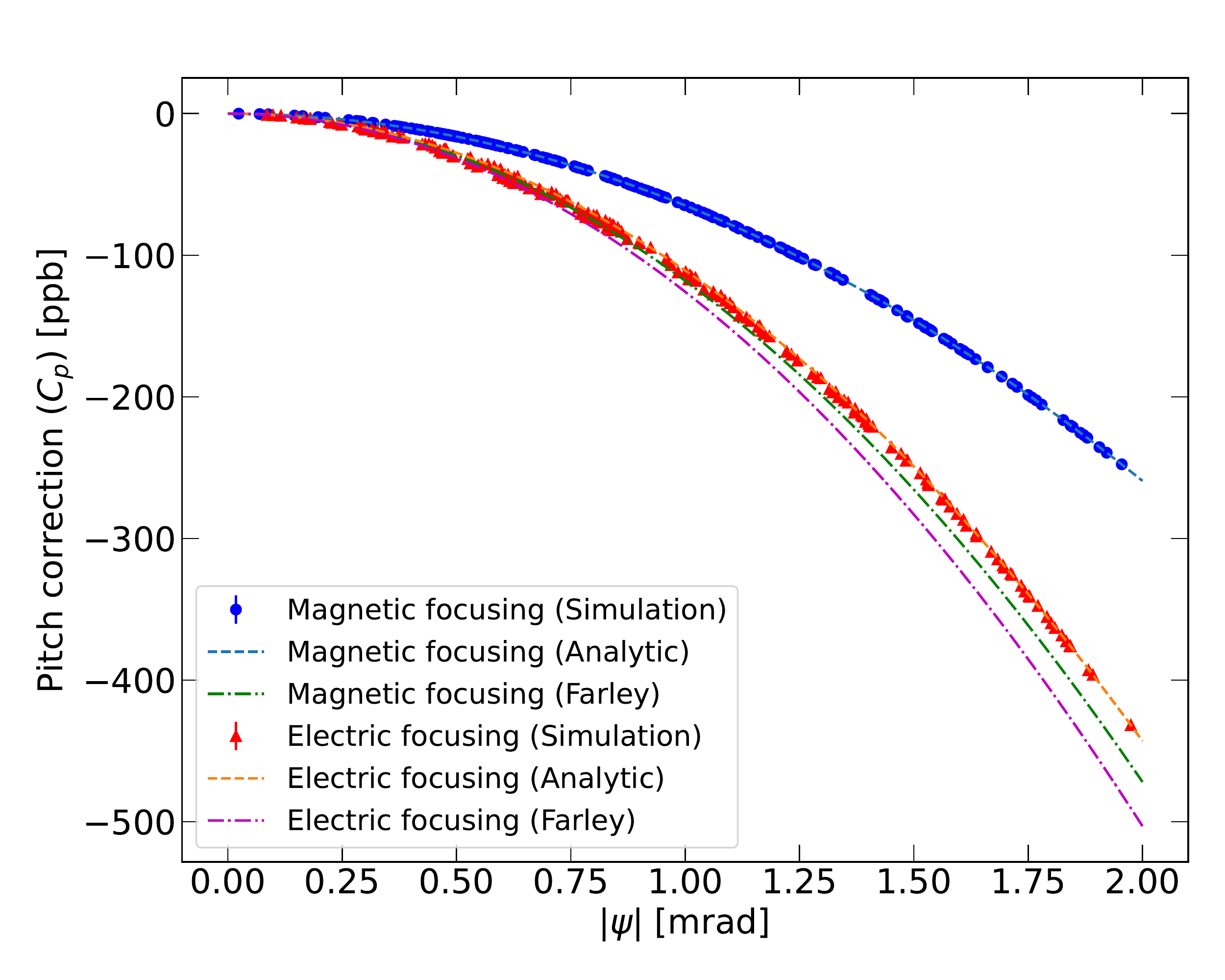}
		\caption{The pitch correction.}
	\end{subfigure}
	\caption{The dispersion correction ($C_d$) and pitch correction ($C_p$) for the proton with a momentum 1 GeV/c, obtained by the high precision spin tracking simulation and the analytic estimation in the magnetic and electric focusing storage rings. Each point was obtained by fitting the simulated spin precession curve for each particle, and agreed well with the analytical estimations, indicated by the dashed lines labelled `Analytic'. Another theoretical prediction for the pitch correction labelled `Farley' from Ref.~\cite{Farley1972} was plotted as reference.}
	\label{fig:Cd_Cp_simulation}
\end{figure*}

The analytic expressions for the dispersion and pitch corrections were again verified with a high precision spin tracking simulation, as shown in Fig.~\ref{fig:Cd_Cp_simulation}. The protons with momentum 1 GeV/c were simulated in 10 m radius magnetic or electric weak focusing storage rings with the field index $n = 0.1$. Each point in the figure represents the dispersion or pitch correction obtained by fitting the longitudinal spin component $S_s$, and the dashed curves are the analytic expressions. One can find there is excellent agreement between the simulation and analytic expressions obtained in the present work.

For the pitch correction, our results were actually rather different from F. Farley's results published in 1972\cite{Farley1972}. He pioneered the pitch correction estimation through a concise set of equations. What he obtained, written in the notations of this paper, is as follows.
\begin{align}
	C_p^{B\text{-focusing}} \text{ (Farley)} &= -\frac{\psi_0^2}{4} \left( 1 + \frac{2G\gamma^2 \omega_y^2 + \omega_{a0}^2}{\gamma^2 (\omega_y^2 - \omega_{a0}^2)} \right), \\
	C_p^{E\text{-focusing}} \text{ (Farley)} &= -\frac{\psi_0^2}{4} \left( 1 + \frac{G^2 (\gamma^2 - 1)^2 \omega_y^2}{\gamma^2 (\omega_y^2 - \omega_{a0}^2)} - \frac{1}{\gamma^2} \right),
\end{align}
where $C_p^{B\text{-focusing}}$ and $C_p^{E\text{-focusing}}$ are the pitch corrections for the magnetic and electric focusing, respectively. There are some missing terms compared to the results of the present work, which undoubtedly caused manifest deviations from the simulation results as shown in Fig.~\ref{fig:Cd_Cp_simulation} (b). We contemplate that the difference between our results and Farley's mainly can be traced to whether the effect from the horizontal equilibrium orbit and the small oscillation of $S_y$ are taken into account. Nonetheless, for the magic momentum particle, his result for the electric focusing becomes identical to our result in Eq.~\eqref{eq:Cp_EF_magic}.

The pitch correction used in the muon $g-2$ experiments at BNL and Fermilab was $C_p = -\langle \psi_0^2 \rangle/4$, without the second term in the parenthesis in Eq.~\eqref{eq:Cp_EF_magic}\cite{Abi2021, Albahri2021_PRAB}. Restoring the second term, however, will not change the result since it is too marginal: $\Delta C_p \sim \mathcal{O} \left( 10^{-10} \right)$.

\section{Discussion} \label{sec:discussion}
In this section, we would like to discuss our results from various aspects to understand more details.

	\subsection{Discrete Focusing Fields}
Storage rings generally consist of discrete focusing elements rather than continuous elements, for many technical reasons. For instance, the electric focusing elements in the muon $g-2$ storage ring in Fermilab, referred to as the electrostatic quadrupoles (ESQs), cover only 43\% of the ring circumference despite the highly symmetric circular design of the lattice\cite{Bennett2006, Abi2021}. Nonetheless, `smoothing' the discrete focusing to the entire circumference works sufficiently well as a simple approximation to roughly estimate the beam parameters. If next order analytical estimations are required, then the transfer matrix method provides a good way to obtain them.

For example, the Hill's equation for the horizontal motion (assuming no dispersion) is given as
\begin{align}
	x''(s) + k(s) x(s) = 0,
\end{align}
where $k(s)$ is the spring-constant-like lattice coefficient as a function of the longitudinal position. One can solve this equation piecewise with the transfer matrix $M(s_0 \rightarrow s)$ with respect to $(x(s), x'(s))$, which effectively gives the betatron tune. Extending this into the case with the momentum offset $\delta$, one can solve the following equation
\begin{align}
	\eta''(s) + k(s, \delta) \eta(s) = 0,
\end{align}
where $\eta(s) \equiv x(s) - D(s) \delta$ is the horizontal betatron oscillation with respect to the equilibrium which is defined by the dispersion function $D(s)$. $k(s, \delta)$ is a modified coefficient that is extended from Eq.~\eqref{eq:eta_MF}, say for instance, in the magnetic focusing case. Then one can obtain more accurate analytical estimation of the betatron tunes and the chromaticities in the discrete focusing storage rings.

	\subsection{Suppression of the Dispersion Correction}
\begin{figure*}[t]
	\centering
	\begin{subfigure}[t]{0.46\textwidth}
		\includegraphics[width=\textwidth]{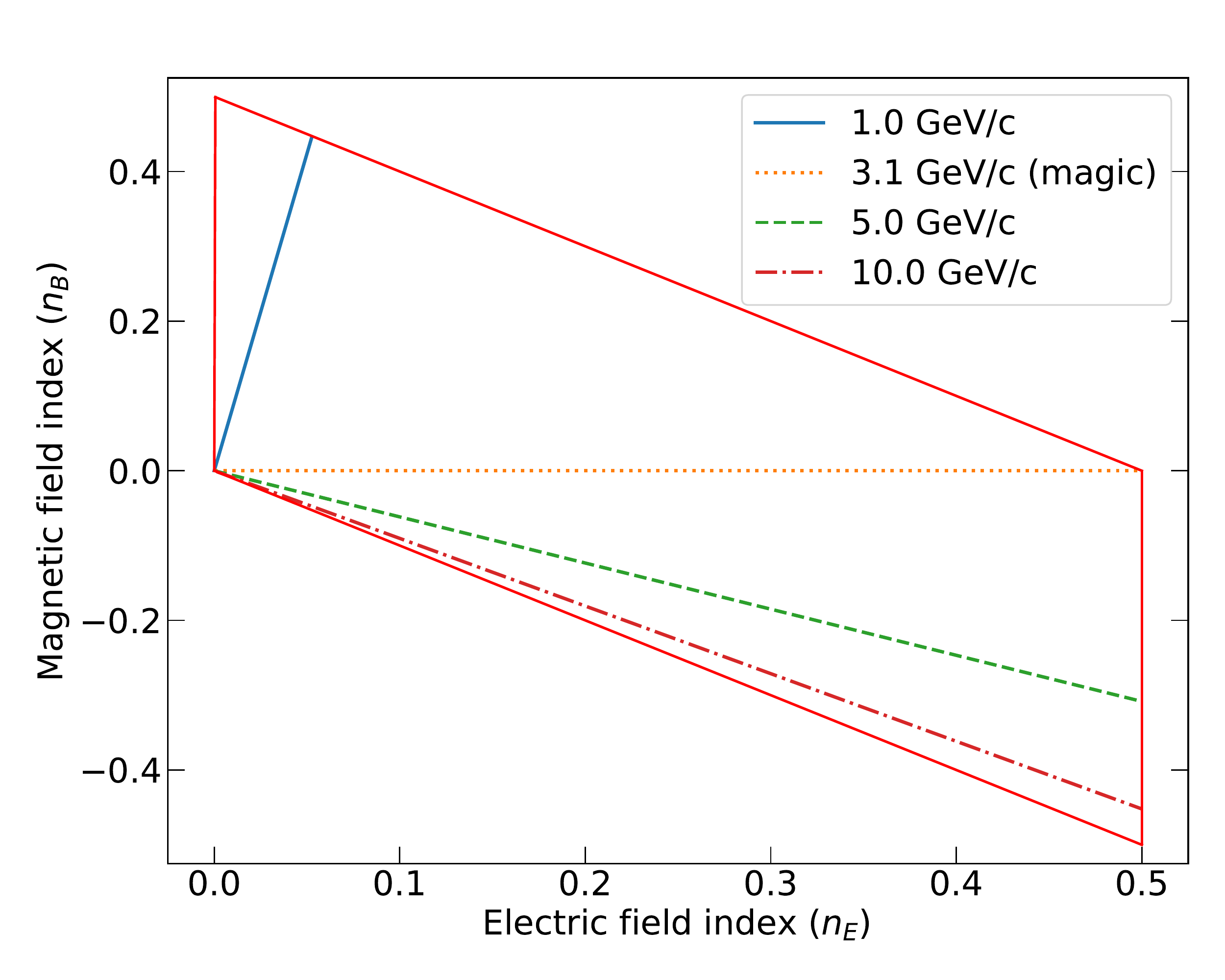}
		\caption{Muon.}
	\end{subfigure}
	~
	\begin{subfigure}[t]{0.46\textwidth}
		\includegraphics[width=\textwidth]{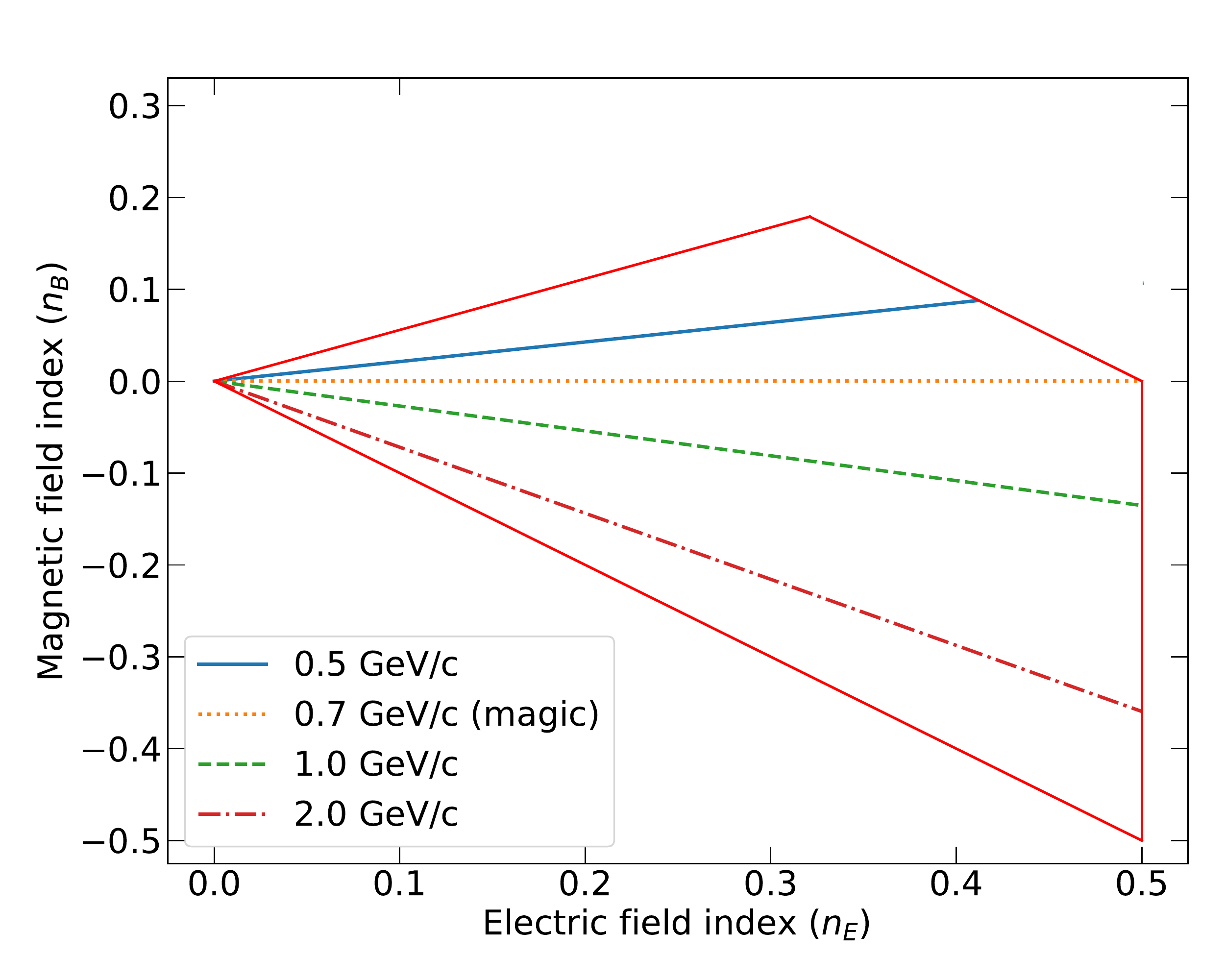}
		\caption{Proton.}
	\end{subfigure}
	\caption{The relation between the magnetic and electric field indices ($n_B$ and $n_E$) to cancel out the dispersion correction for given momenta of the muon and the proton. The field indices must lie within a region enclosed by the red solid lines, to obtain the net vertical focusing and positive momentum. The magic momentum case for each particle is shown for validation, where it needs only the electric focusing to be present ($n_B = 0$).}
	\label{fig:Cd0_CombinedMagEle}
\end{figure*}

We have observed in the previous section, assuming the order of momentum dispersion is about similar to the order of pitch angle, that the dispersion correction naturally dominates over the pitch correction, except in the specific condition that it vanishes. Therefore, in a high precision spin precession experiment such as the muon $g-2$ experiment, it is important to suppress the dispersion correction, and that is precisely the reason why the concept of the magic momentum was developed in one of the earlier muon $g-2$ experiments in CERN\cite{Bailey1977, Bailey1979}. We propose a similar way to suppress the dispersion correction but with the combined magnetic and electric weak focusing, which no longer constrains the momentum to be the magic momentum.

The field index of the combined magnetic and electric focusing will be given as $n_B + n_E$ where $n_B$ and $n_E$ are the field indices for  each focusing element, respectively. The dispersion correction is then approximately given as
\begin{align}
	C_d = -\frac{n_B \delta}{1 - n_B - n_E} - \frac{\beta^2}{G} \left( G - \frac{1}{\gamma^2 - 1} \right) \frac{n_E \delta}{1 - n_B - n_E},
\end{align}
which is nothing but the linear summation for the magnetic and electric focusing cases. Note that the net field index $n_B+n_E$ appears in the denominator, because it is coming from the radial equilibrium orbit, which is determined by the net focusing effect on average. It should satisfy the normal condition $0 < n_B + n_E < 0.5$, to have an effective net vertical focusing, whereas individual $n_B$ or $n_E$ could be negative, indicating vertical defocusing: $|n_B| < 0.5$ and $|n_E| < 0.5$. A specific relation $n_B$ and $n_E$ must satisfy for the dispersion correction to vanish is given as:
\begin{align} \label{eq:nB_nE_relation}
	\frac{n_B}{n_E} = -\frac{\beta^2}{G} \left( G - \frac{1}{\gamma^2 - 1} \right).
\end{align}

The above condition is not quite restrictive on the momentum of the particle. The $n_B/n_E$ ratios for specific momenta of the muon and the proton are shown in Fig.~\ref{fig:Cd0_CombinedMagEle}. Each colored line indicates the above relation between the field indices for a given momentum to cancel out the dispersion correction. The constraints that we discussed require that these field indices $(n_B, n_E)$ must lie within a region enclosed by the red solid lines, otherwise one ends up having no net vertical focusing or non-physical momentum. The momentum $p$ that satisfies Eq.~\eqref{eq:nB_nE_relation} with a given ratio $R \equiv n_B/n_E$, is given as $p/mc = \sqrt{(1 - G R)/G(1 + R)}$. The momentum values are arbitrarily chosen to give an idea of the momentum dependency, but each figure involves a magic momentum for validation, where it should require no magnetic focusing ($n_B = 0$) but only the electric focusing.

\begin{figure}[htb]
	\centering
	\includegraphics[width=0.45\textwidth]{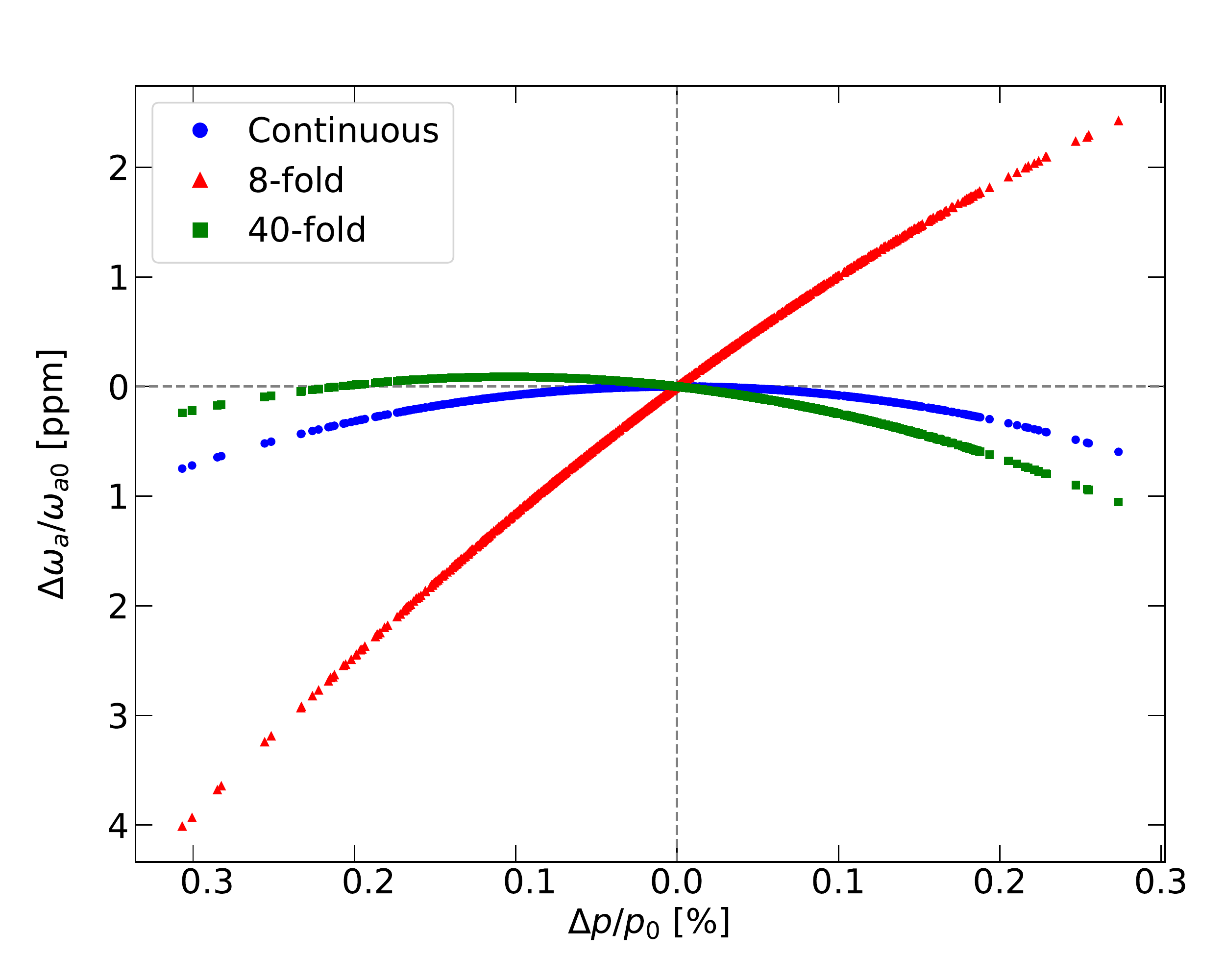}
	\caption{The spin precession frequency relative to the reference, $\Delta \omega_a/\omega_{a0}$, in the combined magnetic and electric focusing storage ring with 5 GeV/c muons from numerical simulations. The magnetic and electric field indices were given such that the dispersion correction $C_d$ is suppressed. Different colors represent different lattice configurations for the focusing elements: (blue) continuous focusing elements around the ring, (red) discrete 8-fold and (green) 40-fold alternating elements.}
	\label{fig:Cd0_CombinedMagEle_simulation}
\end{figure}

The validity of Eq.~\eqref{eq:nB_nE_relation} was tested by numerical simulations as shown in Fig.~\ref{fig:Cd0_CombinedMagEle_simulation}. The spin tracking simulation with 5 GeV/c muons with the momentum spread $0.1\%$ was conducted in the combined magnetic and electric focusing storage ring. Equation~\eqref{eq:nB_nE_relation} gives $n_B/n_E \simeq -0.616$, and therefore the field indices were set to $n_E = 0.1$ and $n_B \simeq -0.062$. In Fig.~\ref{fig:Cd0_CombinedMagEle_simulation}, there are three different plots with different lattice configurations for the focusing elements. Blue points are where they are continuous around the ring, preserving the perfect azimuthal symmetry. Then each magnetic and electric focusing element became localized, alternatingly placed, and made up 8- and 40-fold lattice in red and green points, respectively. Typically, one would have an order of 10 or 100 ppm of the dispersion correction with the given momentum spread. However, thanks to the combined focusing, it was suppressed to sub-ppm level for the azimuthally continuous focusing, where the parabolic pattern clearly indicates the dominant error is from the second-order effect. The dispersion correction gets larger, from continuous to 8-fold lattice, but it does not deviate more than a few ppm level. For the 8-fold symmetry, the effect from discrete elements increased and the linear correction fails to explain the simulation result. Although more realistic numerical simulation with field errors or fringe fields is necessary for solid design of the optimum lattice, we emphasize the aim of this paper is to show the proof-of-principle.

Moreover, practically not all field indices, in either a single or combined focusing lattice, are selectable. Normally they should be determined and tuned very carefully to avoid the low order betatron tune resonances and spin resonances. Nonetheless, this idea of combining the magnetic and electric focusing elements to suppress the dispersion correction can be beneficial, especially for a potential future muon $g-2$ experiment, to improve the statistical sensitivity.

The statistical sensitivity of the $g-2$ frequency in the muon $g-2$ experiments was given as $\sigma_\omega/\omega_{a} \propto 1/(\omega_{a} \gamma \tau_\mu)$\cite{Bennett2007}. For instance, if one stores a 10 GeV/c muon beam in the combined magnetic and electric focusing storage ring, which is roughly 3.3 times higher momentum than the magic momentum which will require roughly 3.3 times stronger magnetic field assuming the same ring radius, then we gain an order of magnitude reduction in the statistical uncertainty for the same number of decay positrons. Of course, this potential new experiment should overcome technical difficulties such as measuring the field gradients to ppb-level precision and having well-controlled fringe fields, et cetera. This needs a further dedicated study.

\section{Conclusion}
In this paper, we derived highly accurate analytical approximations to the beam and spin dynamics of a particle in weak focusing magnetic storage rings, mainly focusing on the beam transverse chromaticities and frequency corrections to the spin precession. Our results are in excellent agreement with high precision spin tracking simulations, and we discussed a new idea of eliminating the dispersion correction with combined magnetic and electric focusing elements.

Our results not only provide solid references to high precision particle physics experiments requiring a deep understanding of the beam or spin dynamics, such as muon $g-2$ experiments or proton/deuteron storage ring EDM experiments, but pave the way to analytically estimate next order approximations to the beam or spin parameters in various storage ring lattice designs.

\acknowledgements
This work was supported by IBS-R017-D1-2021-a00.

\bibliography{weak_focusing_beam_and_spin}

\begin{thebibliography}{40}%
\makeatletter
\providecommand \@ifxundefined [1]{%
 \@ifx{#1\undefined}
}%
\providecommand \@ifnum [1]{%
 \ifnum #1\expandafter \@firstoftwo
 \else \expandafter \@secondoftwo
 \fi
}%
\providecommand \@ifx [1]{%
 \ifx #1\expandafter \@firstoftwo
 \else \expandafter \@secondoftwo
 \fi
}%
\providecommand \natexlab [1]{#1}%
\providecommand \enquote  [1]{``#1''}%
\providecommand \bibnamefont  [1]{#1}%
\providecommand \bibfnamefont [1]{#1}%
\providecommand \citenamefont [1]{#1}%
\providecommand \href@noop [0]{\@secondoftwo}%
\providecommand \href [0]{\begingroup \@sanitize@url \@href}%
\providecommand \@href[1]{\@@startlink{#1}\@@href}%
\providecommand \@@href[1]{\endgroup#1\@@endlink}%
\providecommand \@sanitize@url [0]{\catcode `\\12\catcode `\$12\catcode
  `\&12\catcode `\#12\catcode `\^12\catcode `\_12\catcode `\%12\relax}%
\providecommand \@@startlink[1]{}%
\providecommand \@@endlink[0]{}%
\providecommand \url  [0]{\begingroup\@sanitize@url \@url }%
\providecommand \@url [1]{\endgroup\@href {#1}{\urlprefix }}%
\providecommand \urlprefix  [0]{URL }%
\providecommand \Eprint [0]{\href }%
\providecommand \doibase [0]{https://doi.org/}%
\providecommand \selectlanguage [0]{\@gobble}%
\providecommand \bibinfo  [0]{\@secondoftwo}%
\providecommand \bibfield  [0]{\@secondoftwo}%
\providecommand \translation [1]{[#1]}%
\providecommand \BibitemOpen [0]{}%
\providecommand \bibitemStop [0]{}%
\providecommand \bibitemNoStop [0]{.\EOS\space}%
\providecommand \EOS [0]{\spacefactor3000\relax}%
\providecommand \BibitemShut  [1]{\csname bibitem#1\endcsname}%
\let\auto@bib@innerbib\@empty
%</preamble>
\bibitem [{\citenamefont {Weng}\ and\ \citenamefont {Mane}(1992)}]{Weng1992}%
  \BibitemOpen
  \bibfield  {author} {\bibinfo {author} {\bibfnamefont {W.~T.}\ \bibnamefont
  {Weng}}\ and\ \bibinfo {author} {\bibfnamefont {S.~R.}\ \bibnamefont
  {Mane}},\ }\bibfield  {title} {\bibinfo {title} {Fundamentals of particle
  beam dynamics and phase space},\ }\href {https://doi.org/10.1063/1.42001}
  {\bibfield  {journal} {\bibinfo  {journal} {AIP Conference Proceedings}\
  }\textbf {\bibinfo {volume} {249}},\ \bibinfo {pages} {3} (\bibinfo {year}
  {1992})},\ \Eprint
  {https://arxiv.org/abs/https://aip.scitation.org/doi/pdf/10.1063/1.42001}
  {https://aip.scitation.org/doi/pdf/10.1063/1.42001} \BibitemShut {NoStop}%
\bibitem [{\citenamefont {Edwards}\ and\ \citenamefont
  {Syphers}(2008)}]{Edwards2008}%
  \BibitemOpen
  \bibfield  {author} {\bibinfo {author} {\bibfnamefont {D.~A.}\ \bibnamefont
  {Edwards}}\ and\ \bibinfo {author} {\bibfnamefont {M.~J.}\ \bibnamefont
  {Syphers}},\ }\bibinfo {title} {Transverse linear motion},\ in\ \href
  {https://doi.org/10.1002/9783527617272.ch3} {\emph {\bibinfo {booktitle} {An
  Introduction to the Physics of High Energy Accelerators}}}\ (\bibinfo
  {publisher} {John Wiley \& Sons, Ltd},\ \bibinfo {year} {2008})\BibitemShut
  {NoStop}%
\bibitem [{\citenamefont {Conte}\ and\ \citenamefont
  {MacKay}(2008)}]{Conte2008}%
  \BibitemOpen
  \bibfield  {author} {\bibinfo {author} {\bibfnamefont {M.}~\bibnamefont
  {Conte}}\ and\ \bibinfo {author} {\bibfnamefont {W.~W.}\ \bibnamefont
  {MacKay}},\ }\href {https://doi.org/10.1142/6683} {\emph {\bibinfo {title}
  {An Introduction to the Physics of Particle Accelerators}}},\ \bibinfo
  {edition} {2nd}\ ed.\ (\bibinfo  {publisher} {WORLD SCIENTIFIC},\ \bibinfo
  {year} {2008})\BibitemShut {NoStop}%
\bibitem [{\citenamefont {Martin~Berz}(2015)}]{Berz2015}%
  \BibitemOpen
  \bibfield  {author} {\bibinfo {author} {\bibfnamefont {W.~W.}\ \bibnamefont
  {Martin~Berz}, \bibfnamefont {Kyoko~Makino}},\ }\href
  {https://doi.org/10.1201/b12074} {\emph {\bibinfo {title} {An Introduction to
  Beam Physics}}},\ \bibinfo {edition} {1st}\ ed.\ (\bibinfo  {publisher} {CRC
  Press},\ \bibinfo {year} {2015})\BibitemShut {NoStop}%
\bibitem [{\citenamefont {Lee}(2019)}]{Lee2019}%
  \BibitemOpen
  \bibfield  {author} {\bibinfo {author} {\bibfnamefont {S.~Y.}\ \bibnamefont
  {Lee}},\ }\href {https://doi.org/10.1142/11111} {\emph {\bibinfo {title}
  {Accelerator Physics}}},\ \bibinfo {edition} {4th}\ ed.\ (\bibinfo
  {publisher} {WORLD SCIENTIFIC},\ \bibinfo {year} {2019})\ \Eprint
  {https://arxiv.org/abs/https://www.worldscientific.com/doi/pdf/10.1142/11111}
  {https://www.worldscientific.com/doi/pdf/10.1142/11111} \BibitemShut
  {NoStop}%
\bibitem [{\citenamefont {Bennett}\ \emph {et~al.}(2006)\citenamefont {Bennett}
  \emph {et~al.}}]{Bennett2006}%
  \BibitemOpen
  \bibfield  {author} {\bibinfo {author} {\bibfnamefont {G.~W.}\ \bibnamefont
  {Bennett}} \emph {et~al.} (\bibinfo {collaboration} {Muon g-2
  Collaboration}),\ }\bibfield  {title} {\bibinfo {title} {Final report of the
  e821 muon anomalous magnetic moment measurement at bnl},\ }\href
  {https://doi.org/10.1103/PhysRevD.73.072003} {\bibfield  {journal} {\bibinfo
  {journal} {Phys. Rev. D}\ }\textbf {\bibinfo {volume} {73}},\ \bibinfo
  {pages} {072003} (\bibinfo {year} {2006})}\BibitemShut {NoStop}%
\bibitem [{\citenamefont {Abi}\ \emph {et~al.}(2021)\citenamefont {Abi} \emph
  {et~al.}}]{Abi2021}%
  \BibitemOpen
  \bibfield  {author} {\bibinfo {author} {\bibfnamefont {B.}~\bibnamefont
  {Abi}} \emph {et~al.} (\bibinfo {collaboration} {Muon $g\ensuremath{-}2$
  Collaboration}),\ }\bibfield  {title} {\bibinfo {title} {Measurement of the
  positive muon anomalous magnetic moment to 0.46 ppm},\ }\href
  {https://doi.org/10.1103/PhysRevLett.126.141801} {\bibfield  {journal}
  {\bibinfo  {journal} {Phys. Rev. Lett.}\ }\textbf {\bibinfo {volume} {126}},\
  \bibinfo {pages} {141801} (\bibinfo {year} {2021})}\BibitemShut {NoStop}%
\bibitem [{\citenamefont {Abe}\ \emph {et~al.}(2019)\citenamefont {Abe} \emph
  {et~al.}}]{Abe2019}%
  \BibitemOpen
  \bibfield  {author} {\bibinfo {author} {\bibfnamefont {M.}~\bibnamefont
  {Abe}} \emph {et~al.},\ }\bibfield  {title} {\bibinfo {title} {{A new
  approach for measuring the muon anomalous magnetic moment and electric dipole
  moment}},\ }\bibfield  {journal} {\bibinfo  {journal} {Progress of
  Theoretical and Experimental Physics}\ }\textbf {\bibinfo {volume} {2019}},\
  \href {https://doi.org/10.1093/ptep/ptz030} {10.1093/ptep/ptz030} (\bibinfo
  {year} {2019}),\ \bibinfo {note} {053C02},\ \Eprint
  {https://arxiv.org/abs/https://academic.oup.com/ptep/article-pdf/2019/5/053C02/28746337/ptz030.pdf}
  {https://academic.oup.com/ptep/article-pdf/2019/5/053C02/28746337/ptz030.pdf}
  \BibitemShut {NoStop}%
\bibitem [{\citenamefont {Anastassopoulos}\ \emph {et~al.}(2016)\citenamefont
  {Anastassopoulos} \emph {et~al.}}]{Anastassopoulos2016}%
  \BibitemOpen
  \bibfield  {author} {\bibinfo {author} {\bibfnamefont {V.}~\bibnamefont
  {Anastassopoulos}} \emph {et~al.},\ }\bibfield  {title} {\bibinfo {title} {A
  storage ring experiment to detect a proton electric dipole moment},\ }\href
  {https://doi.org/10.1063/1.4967465} {\bibfield  {journal} {\bibinfo
  {journal} {Review of Scientific Instruments}\ }\textbf {\bibinfo {volume}
  {87}},\ \bibinfo {pages} {115116} (\bibinfo {year} {2016})},\ \Eprint
  {https://arxiv.org/abs/https://aip.scitation.org/doi/pdf/10.1063/1.4967465}
  {https://aip.scitation.org/doi/pdf/10.1063/1.4967465} \BibitemShut {NoStop}%
\bibitem [{\citenamefont {Omarov}\ \emph {et~al.}(2021)\citenamefont {Omarov},
  \citenamefont {Davoudiasl}, \citenamefont {Haciomeroglu}, \citenamefont
  {Lebedev}, \citenamefont {Morse}, \citenamefont {Semertzidis}, \citenamefont
  {Silenko}, \citenamefont {Stephenson},\ and\ \citenamefont
  {Suleiman}}]{Omarov2021}%
  \BibitemOpen
  \bibfield  {author} {\bibinfo {author} {\bibfnamefont {Z.}~\bibnamefont
  {Omarov}}, \bibinfo {author} {\bibfnamefont {H.}~\bibnamefont {Davoudiasl}},
  \bibinfo {author} {\bibfnamefont {S.}~\bibnamefont {Haciomeroglu}}, \bibinfo
  {author} {\bibfnamefont {V.}~\bibnamefont {Lebedev}}, \bibinfo {author}
  {\bibfnamefont {W.~M.}\ \bibnamefont {Morse}}, \bibinfo {author}
  {\bibfnamefont {Y.~K.}\ \bibnamefont {Semertzidis}}, \bibinfo {author}
  {\bibfnamefont {A.~J.}\ \bibnamefont {Silenko}}, \bibinfo {author}
  {\bibfnamefont {E.~J.}\ \bibnamefont {Stephenson}},\ and\ \bibinfo {author}
  {\bibfnamefont {R.}~\bibnamefont {Suleiman}},\ }\href@noop {} {\bibinfo
  {title} {Comprehensive symmetric-hybrid ring design for pedm experiment at
  below $10^{-29}e\cdot$cm}} (\bibinfo {year} {2021}),\ \Eprint
  {https://arxiv.org/abs/2007.10332} {arXiv:2007.10332 [physics.acc-ph]}
  \BibitemShut {NoStop}%
\bibitem [{\citenamefont {Eversmann}\ \emph {et~al.}(2015)\citenamefont
  {Eversmann} \emph {et~al.}}]{Eversmann2015}%
  \BibitemOpen
  \bibfield  {author} {\bibinfo {author} {\bibfnamefont {D.}~\bibnamefont
  {Eversmann}} \emph {et~al.} (\bibinfo {collaboration} {JEDI collaboration}),\
  }\bibfield  {title} {\bibinfo {title} {New method for a continuous
  determination of the spin tune in storage rings and implications for
  precision experiments},\ }\href
  {https://doi.org/10.1103/PhysRevLett.115.094801} {\bibfield  {journal}
  {\bibinfo  {journal} {Phys. Rev. Lett.}\ }\textbf {\bibinfo {volume} {115}},\
  \bibinfo {pages} {094801} (\bibinfo {year} {2015})}\BibitemShut {NoStop}%
\bibitem [{\citenamefont {Guidoboni}\ \emph {et~al.}(2016)\citenamefont
  {Guidoboni} \emph {et~al.}}]{Guidoboni2016}%
  \BibitemOpen
  \bibfield  {author} {\bibinfo {author} {\bibfnamefont {G.}~\bibnamefont
  {Guidoboni}} \emph {et~al.} (\bibinfo {collaboration} {JEDI Collaboration}),\
  }\bibfield  {title} {\bibinfo {title} {How to reach a thousand-second
  in-plane polarization lifetime with $0.97\text{\ensuremath{-}}\mathrm{GeV}/c$
  deuterons in a storage ring},\ }\href
  {https://doi.org/10.1103/PhysRevLett.117.054801} {\bibfield  {journal}
  {\bibinfo  {journal} {Phys. Rev. Lett.}\ }\textbf {\bibinfo {volume} {117}},\
  \bibinfo {pages} {054801} (\bibinfo {year} {2016})}\BibitemShut {NoStop}%
\bibitem [{\citenamefont {Hempelmann}\ \emph {et~al.}(2017)\citenamefont
  {Hempelmann} \emph {et~al.}}]{Hempelmann2017}%
  \BibitemOpen
  \bibfield  {author} {\bibinfo {author} {\bibfnamefont {N.}~\bibnamefont
  {Hempelmann}} \emph {et~al.} (\bibinfo {collaboration} {JEDI
  Collaboration}),\ }\bibfield  {title} {\bibinfo {title} {Phase locking the
  spin precession in a storage ring},\ }\href
  {https://doi.org/10.1103/PhysRevLett.119.014801} {\bibfield  {journal}
  {\bibinfo  {journal} {Phys. Rev. Lett.}\ }\textbf {\bibinfo {volume} {119}},\
  \bibinfo {pages} {014801} (\bibinfo {year} {2017})}\BibitemShut {NoStop}%
\bibitem [{\citenamefont {Saleev}\ \emph {et~al.}(2017)\citenamefont {Saleev}
  \emph {et~al.}}]{Saleev2017}%
  \BibitemOpen
  \bibfield  {author} {\bibinfo {author} {\bibfnamefont {A.}~\bibnamefont
  {Saleev}} \emph {et~al.} (\bibinfo {collaboration} {JEDI collaboration}),\
  }\bibfield  {title} {\bibinfo {title} {Spin tune mapping as a novel tool to
  probe the spin dynamics in storage rings},\ }\href
  {https://doi.org/10.1103/PhysRevAccelBeams.20.072801} {\bibfield  {journal}
  {\bibinfo  {journal} {Phys. Rev. Accel. Beams}\ }\textbf {\bibinfo {volume}
  {20}},\ \bibinfo {pages} {072801} (\bibinfo {year} {2017})}\BibitemShut
  {NoStop}%
\bibitem [{\citenamefont {Chang}\ \emph {et~al.}(2019)\citenamefont {Chang},
  \citenamefont {Hac\ifmmode \imath \else \i
  \fi{}\"omero\ifmmode~\breve{g}\else \u{g}\fi{}lu}, \citenamefont {Kim},
  \citenamefont {Lee}, \citenamefont {Park},\ and\ \citenamefont
  {Semertzidis}}]{Chang2019}%
  \BibitemOpen
  \bibfield  {author} {\bibinfo {author} {\bibfnamefont {S.~P.}\ \bibnamefont
  {Chang}}, \bibinfo {author} {\bibfnamefont {S.}~\bibnamefont {Hac\ifmmode
  \imath \else \i \fi{}\"omero\ifmmode~\breve{g}\else \u{g}\fi{}lu}}, \bibinfo
  {author} {\bibfnamefont {O.}~\bibnamefont {Kim}}, \bibinfo {author}
  {\bibfnamefont {S.}~\bibnamefont {Lee}}, \bibinfo {author} {\bibfnamefont
  {S.}~\bibnamefont {Park}},\ and\ \bibinfo {author} {\bibfnamefont {Y.~K.}\
  \bibnamefont {Semertzidis}},\ }\bibfield  {title} {\bibinfo {title}
  {Axionlike dark matter search using the storage ring edm method},\ }\href
  {https://doi.org/10.1103/PhysRevD.99.083002} {\bibfield  {journal} {\bibinfo
  {journal} {Phys. Rev. D}\ }\textbf {\bibinfo {volume} {99}},\ \bibinfo
  {pages} {083002} (\bibinfo {year} {2019})}\BibitemShut {NoStop}%
\bibitem [{\citenamefont {Graham}\ \emph {et~al.}(2021)\citenamefont {Graham},
  \citenamefont {Hac\ifmmode \imath \else \i
  \fi{}\"omero\ifmmode~\breve{g}\else \u{g}\fi{}lu}, \citenamefont {Kaplan},
  \citenamefont {Omarov}, \citenamefont {Rajendran},\ and\ \citenamefont
  {Semertzidis}}]{Graham2021}%
  \BibitemOpen
  \bibfield  {author} {\bibinfo {author} {\bibfnamefont {P.~W.}\ \bibnamefont
  {Graham}}, \bibinfo {author} {\bibfnamefont {S.}~\bibnamefont {Hac\ifmmode
  \imath \else \i \fi{}\"omero\ifmmode~\breve{g}\else \u{g}\fi{}lu}}, \bibinfo
  {author} {\bibfnamefont {D.~E.}\ \bibnamefont {Kaplan}}, \bibinfo {author}
  {\bibfnamefont {Z.}~\bibnamefont {Omarov}}, \bibinfo {author} {\bibfnamefont
  {S.}~\bibnamefont {Rajendran}},\ and\ \bibinfo {author} {\bibfnamefont
  {Y.~K.}\ \bibnamefont {Semertzidis}},\ }\bibfield  {title} {\bibinfo {title}
  {Storage ring probes of dark matter and dark energy},\ }\href
  {https://doi.org/10.1103/PhysRevD.103.055010} {\bibfield  {journal} {\bibinfo
   {journal} {Phys. Rev. D}\ }\textbf {\bibinfo {volume} {103}},\ \bibinfo
  {pages} {055010} (\bibinfo {year} {2021})}\BibitemShut {NoStop}%
\bibitem [{\citenamefont {Kim}\ and\ \citenamefont
  {Semertzidis}(2021)}]{Kim2021}%
  \BibitemOpen
  \bibfield  {author} {\bibinfo {author} {\bibfnamefont {O.}~\bibnamefont
  {Kim}}\ and\ \bibinfo {author} {\bibfnamefont {Y.~K.}\ \bibnamefont
  {Semertzidis}},\ }\bibfield  {title} {\bibinfo {title} {New method of probing
  an oscillating edm induced by axionlike dark matter using an rf wien filter
  in storage rings},\ }\href {https://doi.org/10.1103/PhysRevD.104.096006}
  {\bibfield  {journal} {\bibinfo  {journal} {Phys. Rev. D}\ }\textbf {\bibinfo
  {volume} {104}},\ \bibinfo {pages} {096006} (\bibinfo {year}
  {2021})}\BibitemShut {NoStop}%
\bibitem [{\citenamefont {Bailey}\ \emph {et~al.}(1979)\citenamefont {Bailey}
  \emph {et~al.}}]{Bailey1979}%
  \BibitemOpen
  \bibfield  {author} {\bibinfo {author} {\bibfnamefont {J.}~\bibnamefont
  {Bailey}} \emph {et~al.},\ }\bibfield  {title} {\bibinfo {title} {Final
  report on the cern muon storage ring including the anomalous magnetic moment
  and the electric dipole moment of the muon, and a direct test of relativistic
  time dilation},\ }\href
  {https://doi.org/https://doi.org/10.1016/0550-3213(79)90292-X} {\bibfield
  {journal} {\bibinfo  {journal} {Nuclear Physics B}\ }\textbf {\bibinfo
  {volume} {150}},\ \bibinfo {pages} {1} (\bibinfo {year} {1979})}\BibitemShut
  {NoStop}%
\bibitem [{\citenamefont {Albahri}\ \emph
  {et~al.}(2021{\natexlab{a}})\citenamefont {Albahri} \emph
  {et~al.}}]{Albahri2021_PRAB}%
  \BibitemOpen
  \bibfield  {author} {\bibinfo {author} {\bibfnamefont {T.}~\bibnamefont
  {Albahri}} \emph {et~al.} (\bibinfo {collaboration} {Muon $g\ensuremath{-}2$
  Collaboration}),\ }\bibfield  {title} {\bibinfo {title} {Beam dynamics
  corrections to the run-1 measurement of the muon anomalous magnetic moment at
  fermilab},\ }\href {https://doi.org/10.1103/PhysRevAccelBeams.24.044002}
  {\bibfield  {journal} {\bibinfo  {journal} {Phys. Rev. Accel. Beams}\
  }\textbf {\bibinfo {volume} {24}},\ \bibinfo {pages} {044002} (\bibinfo
  {year} {2021}{\natexlab{a}})}\BibitemShut {NoStop}%
\bibitem [{\citenamefont {Albahri}\ \emph
  {et~al.}(2021{\natexlab{b}})\citenamefont {Albahri} \emph
  {et~al.}}]{Albahri2021_PRD}%
  \BibitemOpen
  \bibfield  {author} {\bibinfo {author} {\bibfnamefont {T.}~\bibnamefont
  {Albahri}} \emph {et~al.} (\bibinfo {collaboration} {Muon $g\ensuremath{-}2$
  Collaboration}),\ }\bibfield  {title} {\bibinfo {title} {Measurement of the
  anomalous precession frequency of the muon in the fermilab muon
  $g\ensuremath{-}2$ experiment},\ }\href
  {https://doi.org/10.1103/PhysRevD.103.072002} {\bibfield  {journal} {\bibinfo
   {journal} {Phys. Rev. D}\ }\textbf {\bibinfo {volume} {103}},\ \bibinfo
  {pages} {072002} (\bibinfo {year} {2021}{\natexlab{b}})}\BibitemShut
  {NoStop}%
\bibitem [{\citenamefont {Jackson}(1999)}]{Jackson1999}%
  \BibitemOpen
  \bibfield  {author} {\bibinfo {author} {\bibfnamefont {J.}~\bibnamefont
  {Jackson}},\ }\href {http://cdsweb.cern.ch/record/490457} {\emph {\bibinfo
  {title} {Classical electrodynamics}}},\ \bibinfo {edition} {3rd}\ ed.\
  (\bibinfo  {publisher} {Wiley},\ \bibinfo {address} {New York, {NY}},\
  \bibinfo {year} {1999})\BibitemShut {NoStop}%
\bibitem [{\citenamefont {Mane}(2008)}]{Mane2008}%
  \BibitemOpen
  \bibfield  {author} {\bibinfo {author} {\bibfnamefont {S.}~\bibnamefont
  {Mane}},\ }\bibfield  {title} {\bibinfo {title} {Orbital dynamics in a
  storage ring with electrostatic bending},\ }\href
  {https://doi.org/https://doi.org/10.1016/j.nima.2008.08.087} {\bibfield
  {journal} {\bibinfo  {journal} {Nuclear Instruments and Methods in Physics
  Research Section A: Accelerators, Spectrometers, Detectors and Associated
  Equipment}\ }\textbf {\bibinfo {volume} {596}},\ \bibinfo {pages} {288}
  (\bibinfo {year} {2008})}\BibitemShut {NoStop}%
\bibitem [{\citenamefont {Mane}(2012)}]{Mane2012}%
  \BibitemOpen
  \bibfield  {author} {\bibinfo {author} {\bibfnamefont {S.}~\bibnamefont
  {Mane}},\ }\bibfield  {title} {\bibinfo {title} {Orbital and spin motion in a
  storage ring with static electric and magnetic fields},\ }\href
  {https://doi.org/https://doi.org/10.1016/j.nima.2012.05.098} {\bibfield
  {journal} {\bibinfo  {journal} {Nuclear Instruments and Methods in Physics
  Research Section A: Accelerators, Spectrometers, Detectors and Associated
  Equipment}\ }\textbf {\bibinfo {volume} {687}},\ \bibinfo {pages} {40}
  (\bibinfo {year} {2012})}\BibitemShut {NoStop}%
\bibitem [{\citenamefont {Semertzidis}\ \emph {et~al.}(2003)\citenamefont
  {Semertzidis}, \citenamefont {Bennett}, \citenamefont {Efstathiadis},
  \citenamefont {Krienen}, \citenamefont {Larsen}, \citenamefont {Lee},
  \citenamefont {Morse}, \citenamefont {Orlov}, \citenamefont {Ozben},
  \citenamefont {Roberts}, \citenamefont {Snydstrup},\ and\ \citenamefont
  {Warburton}}]{Semertzidis2003}%
  \BibitemOpen
  \bibfield  {author} {\bibinfo {author} {\bibfnamefont {Y.~K.}\ \bibnamefont
  {Semertzidis}}, \bibinfo {author} {\bibfnamefont {G.}~\bibnamefont
  {Bennett}}, \bibinfo {author} {\bibfnamefont {E.}~\bibnamefont
  {Efstathiadis}}, \bibinfo {author} {\bibfnamefont {F.}~\bibnamefont
  {Krienen}}, \bibinfo {author} {\bibfnamefont {R.}~\bibnamefont {Larsen}},
  \bibinfo {author} {\bibfnamefont {Y.}~\bibnamefont {Lee}}, \bibinfo {author}
  {\bibfnamefont {W.~M.}\ \bibnamefont {Morse}}, \bibinfo {author}
  {\bibfnamefont {Y.}~\bibnamefont {Orlov}}, \bibinfo {author} {\bibfnamefont
  {C.~S.}\ \bibnamefont {Ozben}}, \bibinfo {author} {\bibfnamefont
  {B.}~\bibnamefont {Roberts}}, \bibinfo {author} {\bibfnamefont {L.~P.}\
  \bibnamefont {Snydstrup}},\ and\ \bibinfo {author} {\bibfnamefont {D.~S.}\
  \bibnamefont {Warburton}},\ }\bibfield  {title} {\bibinfo {title} {The
  brookhaven muon (g−2) storage ring high voltage quadrupoles},\ }\href
  {https://doi.org/https://doi.org/10.1016/S0168-9002(03)00999-9} {\bibfield
  {journal} {\bibinfo  {journal} {Nuclear Instruments and Methods in Physics
  Research Section A: Accelerators, Spectrometers, Detectors and Associated
  Equipment}\ }\textbf {\bibinfo {volume} {503}},\ \bibinfo {pages} {458}
  (\bibinfo {year} {2003})}\BibitemShut {NoStop}%
\bibitem [{\citenamefont {Runge}(1895)}]{Runge1895}%
  \BibitemOpen
  \bibfield  {author} {\bibinfo {author} {\bibfnamefont {C.}~\bibnamefont
  {Runge}},\ }\bibfield  {title} {\bibinfo {title} {Ueber die numerische
  aufl{\"o}sung von differentialgleichungen},\ }\href
  {https://doi.org/10.1007/BF01446807} {\bibfield  {journal} {\bibinfo
  {journal} {Mathematische Annalen}\ }\textbf {\bibinfo {volume} {46}},\
  \bibinfo {pages} {167} (\bibinfo {year} {1895})}\BibitemShut {NoStop}%
\bibitem [{\citenamefont {Kutta}(1901)}]{Kutta1901}%
  \BibitemOpen
  \bibfield  {author} {\bibinfo {author} {\bibfnamefont {W.}~\bibnamefont
  {Kutta}},\ }\bibfield  {title} {\bibinfo {title} {Beitrag zur
  n\"aherungsweisen {I}ntegration totaler {D}ifferentialgleichungen},\
  }\href@noop {} {\bibfield  {journal} {\bibinfo  {journal} {Zeit. Math.
  Phys.}\ }\textbf {\bibinfo {volume} {46}},\ \bibinfo {pages} {435} (\bibinfo
  {year} {1901})}\BibitemShut {NoStop}%
\bibitem [{\citenamefont {Fehlberg}(1969)}]{Fehlberg1969}%
  \BibitemOpen
  \bibfield  {author} {\bibinfo {author} {\bibfnamefont {E.}~\bibnamefont
  {Fehlberg}},\ }\href {https://ntrs.nasa.gov/citations/19690021375} {\emph
  {\bibinfo {title} {Low-order classical Runge-Kutta formulas with stepsize
  control and their application to some heat transfer problems}}},\ \bibinfo
  {type} {Tech. Rep.}\ \bibinfo {number} {19690021375}\ (\bibinfo
  {institution} {National Aeronautics and Space Administration},\ \bibinfo
  {address} {Washington, D.C.},\ \bibinfo {year} {1969})\BibitemShut {NoStop}%
\bibitem [{\citenamefont {Weisskopf}\ \emph {et~al.}(2019)\citenamefont
  {Weisskopf}, \citenamefont {Tarazona},\ and\ \citenamefont
  {Berz}}]{Weisskopf2019}%
  \BibitemOpen
  \bibfield  {author} {\bibinfo {author} {\bibfnamefont {A.}~\bibnamefont
  {Weisskopf}}, \bibinfo {author} {\bibfnamefont {D.}~\bibnamefont
  {Tarazona}},\ and\ \bibinfo {author} {\bibfnamefont {M.}~\bibnamefont
  {Berz}},\ }\bibfield  {title} {\bibinfo {title} {Computation and consequences
  of high order amplitude- and parameter-dependent tune shifts in storage rings
  for high precision measurements},\ }\href
  {https://doi.org/10.1142/S0217751X19420119} {\bibfield  {journal} {\bibinfo
  {journal} {International Journal of Modern Physics A}\ }\textbf {\bibinfo
  {volume} {34}},\ \bibinfo {pages} {1942011} (\bibinfo {year} {2019})},\
  \Eprint {https://arxiv.org/abs/https://doi.org/10.1142/S0217751X19420119}
  {https://doi.org/10.1142/S0217751X19420119} \BibitemShut {NoStop}%
\bibitem [{\citenamefont {Thomas}(1926)}]{Thomas1926}%
  \BibitemOpen
  \bibfield  {author} {\bibinfo {author} {\bibfnamefont {L.~H.}\ \bibnamefont
  {Thomas}},\ }\bibfield  {title} {\bibinfo {title} {The motion of the spinning
  electron},\ }\href {https://doi.org/10.1038/117514a0} {\bibfield  {journal}
  {\bibinfo  {journal} {Nature}\ }\textbf {\bibinfo {volume} {117}},\ \bibinfo
  {pages} {514} (\bibinfo {year} {1926})}\BibitemShut {NoStop}%
\bibitem [{\citenamefont {Bargmann}\ \emph {et~al.}(1959)\citenamefont
  {Bargmann}, \citenamefont {Michel},\ and\ \citenamefont
  {Telegdi}}]{Bargmann1959}%
  \BibitemOpen
  \bibfield  {author} {\bibinfo {author} {\bibfnamefont {V.}~\bibnamefont
  {Bargmann}}, \bibinfo {author} {\bibfnamefont {L.}~\bibnamefont {Michel}},\
  and\ \bibinfo {author} {\bibfnamefont {V.~L.}\ \bibnamefont {Telegdi}},\
  }\bibfield  {title} {\bibinfo {title} {Precession of the polarization of
  particles moving in a homogeneous electromagnetic field},\ }\href
  {https://doi.org/10.1103/PhysRevLett.2.435} {\bibfield  {journal} {\bibinfo
  {journal} {Phys. Rev. Lett.}\ }\textbf {\bibinfo {volume} {2}},\ \bibinfo
  {pages} {435} (\bibinfo {year} {1959})}\BibitemShut {NoStop}%
\bibitem [{\citenamefont {Sahoo}(2017)}]{Sahoo2017}%
  \BibitemOpen
  \bibfield  {author} {\bibinfo {author} {\bibfnamefont {B.~K.}\ \bibnamefont
  {Sahoo}},\ }\bibfield  {title} {\bibinfo {title} {Improved limits on the
  hadronic and semihadronic $cp$ violating parameters and role of a dark force
  carrier in the electric dipole moment of $^{199}\mathrm{Hg}$},\ }\href
  {https://doi.org/10.1103/PhysRevD.95.013002} {\bibfield  {journal} {\bibinfo
  {journal} {Phys. Rev. D}\ }\textbf {\bibinfo {volume} {95}},\ \bibinfo
  {pages} {013002} (\bibinfo {year} {2017})}\BibitemShut {NoStop}%
\bibitem [{\citenamefont {Bennett}\ \emph {et~al.}(2009)\citenamefont {Bennett}
  \emph {et~al.}}]{Bennett2009}%
  \BibitemOpen
  \bibfield  {author} {\bibinfo {author} {\bibfnamefont {G.~W.}\ \bibnamefont
  {Bennett}} \emph {et~al.} (\bibinfo {collaboration} {Muon (g-2)
  Collaboration}),\ }\bibfield  {title} {\bibinfo {title} {Improved limit on
  the muon electric dipole moment},\ }\href
  {https://doi.org/10.1103/PhysRevD.80.052008} {\bibfield  {journal} {\bibinfo
  {journal} {Phys. Rev. D}\ }\textbf {\bibinfo {volume} {80}},\ \bibinfo
  {pages} {052008} (\bibinfo {year} {2009})}\BibitemShut {NoStop}%
\bibitem [{\citenamefont {Silenko}(2006)}]{Silenko2006}%
  \BibitemOpen
  \bibfield  {author} {\bibinfo {author} {\bibfnamefont {A.~J.}\ \bibnamefont
  {Silenko}},\ }\bibfield  {title} {\bibinfo {title} {Equation of spin motion
  in storage rings in the cylindrical coordinate system},\ }\href
  {https://doi.org/10.1103/PhysRevSTAB.9.034003} {\bibfield  {journal}
  {\bibinfo  {journal} {Phys. Rev. ST Accel. Beams}\ }\textbf {\bibinfo
  {volume} {9}},\ \bibinfo {pages} {034003} (\bibinfo {year}
  {2006})}\BibitemShut {NoStop}%
\bibitem [{\citenamefont {Fukuyama}\ and\ \citenamefont
  {Silenko}(2013)}]{Fukuyama2013}%
  \BibitemOpen
  \bibfield  {author} {\bibinfo {author} {\bibfnamefont {T.}~\bibnamefont
  {Fukuyama}}\ and\ \bibinfo {author} {\bibfnamefont {A.~J.}\ \bibnamefont
  {Silenko}},\ }\bibfield  {title} {\bibinfo {title} {Derivation of generalized
  {T}homas–{B}argmann–{M}ichel–{T}elegdi equation for a particle with
  electric dipole moment},\ }\href {https://doi.org/10.1142/S0217751X13501479}
  {\bibfield  {journal} {\bibinfo  {journal} {International Journal of Modern
  Physics A}\ }\textbf {\bibinfo {volume} {28}},\ \bibinfo {pages} {1350147}
  (\bibinfo {year} {2013})}\BibitemShut {NoStop}%
\bibitem [{\citenamefont {Hoffstaetter}(2002)}]{Hoffstaetter2002}%
  \BibitemOpen
  \bibfield  {author} {\bibinfo {author} {\bibfnamefont {G.}~\bibnamefont
  {Hoffstaetter}},\ }\bibfield  {title} {\bibinfo {title} {{Accelerator rings
  with polarized beams and spin manipulation}},\ }in\ \href@noop {} {\emph
  {\bibinfo {booktitle} {{1st Summer School and Workshop on COSY Physics (CSS
  2002)}}}}\ (\bibinfo {year} {2002})\ pp.\ \bibinfo {pages}
  {85--126}\BibitemShut {NoStop}%
\bibitem [{\citenamefont {Mane}\ \emph {et~al.}(2005)\citenamefont {Mane},
  \citenamefont {Shatunov},\ and\ \citenamefont {Yokoya}}]{Mane2005}%
  \BibitemOpen
  \bibfield  {author} {\bibinfo {author} {\bibfnamefont {S.~R.}\ \bibnamefont
  {Mane}}, \bibinfo {author} {\bibfnamefont {Y.~M.}\ \bibnamefont {Shatunov}},\
  and\ \bibinfo {author} {\bibfnamefont {K.}~\bibnamefont {Yokoya}},\
  }\bibfield  {title} {\bibinfo {title} {Spin-polarized charged particle beams
  in high-energy accelerators},\ }\href
  {https://doi.org/10.1088/0034-4885/68/9/r01} {\bibfield  {journal} {\bibinfo
  {journal} {Reports on Progress in Physics}\ }\textbf {\bibinfo {volume}
  {68}},\ \bibinfo {pages} {1997} (\bibinfo {year} {2005})}\BibitemShut
  {NoStop}%
\bibitem [{\citenamefont {Barber}\ \emph {et~al.}(2001)\citenamefont {Barber},
  \citenamefont {Hoffstätter},\ and\ \citenamefont {Vogt}}]{Barber2001}%
  \BibitemOpen
  \bibfield  {author} {\bibinfo {author} {\bibfnamefont {D.~P.}\ \bibnamefont
  {Barber}}, \bibinfo {author} {\bibfnamefont {G.~H.}\ \bibnamefont
  {Hoffstätter}},\ and\ \bibinfo {author} {\bibfnamefont {M.}~\bibnamefont
  {Vogt}},\ }\bibfield  {title} {\bibinfo {title} {Using the amplitude
  dependent spin tune to study high order spin–orbit resonances in storage
  rings},\ }\href {https://doi.org/10.1063/1.1384158} {\bibfield  {journal}
  {\bibinfo  {journal} {AIP Conference Proceedings}\ }\textbf {\bibinfo
  {volume} {570}},\ \bibinfo {pages} {751} (\bibinfo {year} {2001})},\ \Eprint
  {https://arxiv.org/abs/https://aip.scitation.org/doi/pdf/10.1063/1.1384158}
  {https://aip.scitation.org/doi/pdf/10.1063/1.1384158} \BibitemShut {NoStop}%
\bibitem [{\citenamefont {Bailey}\ \emph {et~al.}(1977)\citenamefont {Bailey},
  \citenamefont {Borer}, \citenamefont {Combley}, \citenamefont {Drumm},
  \citenamefont {Farley}, \citenamefont {Field}, \citenamefont {Flegel},
  \citenamefont {Hattersley}, \citenamefont {Krienen}, \citenamefont {Lange},
  \citenamefont {Picasso},\ and\ \citenamefont {{Von Rüden}}}]{Bailey1977}%
  \BibitemOpen
  \bibfield  {author} {\bibinfo {author} {\bibfnamefont {J.}~\bibnamefont
  {Bailey}}, \bibinfo {author} {\bibfnamefont {K.}~\bibnamefont {Borer}},
  \bibinfo {author} {\bibfnamefont {F.}~\bibnamefont {Combley}}, \bibinfo
  {author} {\bibfnamefont {H.}~\bibnamefont {Drumm}}, \bibinfo {author}
  {\bibfnamefont {F.}~\bibnamefont {Farley}}, \bibinfo {author} {\bibfnamefont
  {J.}~\bibnamefont {Field}}, \bibinfo {author} {\bibfnamefont
  {W.}~\bibnamefont {Flegel}}, \bibinfo {author} {\bibfnamefont
  {P.}~\bibnamefont {Hattersley}}, \bibinfo {author} {\bibfnamefont
  {F.}~\bibnamefont {Krienen}}, \bibinfo {author} {\bibfnamefont
  {F.}~\bibnamefont {Lange}}, \bibinfo {author} {\bibfnamefont
  {E.}~\bibnamefont {Picasso}},\ and\ \bibinfo {author} {\bibfnamefont
  {W.}~\bibnamefont {{Von Rüden}}},\ }\bibfield  {title} {\bibinfo {title}
  {The anomalous magnetic moment of positive and negative muons},\ }\href
  {https://doi.org/https://doi.org/10.1016/0370-2693(77)90199-X} {\bibfield
  {journal} {\bibinfo  {journal} {Physics Letters B}\ }\textbf {\bibinfo
  {volume} {68}},\ \bibinfo {pages} {191} (\bibinfo {year} {1977})}\BibitemShut
  {NoStop}%
\bibitem [{\citenamefont {Farley}(1972)}]{Farley1972}%
  \BibitemOpen
  \bibfield  {author} {\bibinfo {author} {\bibfnamefont {F.}~\bibnamefont
  {Farley}},\ }\bibfield  {title} {\bibinfo {title} {Pitch correction in (g-2)
  experiments},\ }\href
  {https://doi.org/https://doi.org/10.1016/0370-2693(72)90718-6} {\bibfield
  {journal} {\bibinfo  {journal} {Physics Letters B}\ }\textbf {\bibinfo
  {volume} {42}},\ \bibinfo {pages} {66 } (\bibinfo {year} {1972})}\BibitemShut
  {NoStop}%
\bibitem [{\citenamefont {Bennett}\ \emph {et~al.}(2007)\citenamefont {Bennett}
  \emph {et~al.}}]{Bennett2007}%
  \BibitemOpen
  \bibfield  {author} {\bibinfo {author} {\bibfnamefont {G.}~\bibnamefont
  {Bennett}} \emph {et~al.},\ }\bibfield  {title} {\bibinfo {title}
  {Statistical equations and methods applied to the precision muon (g-2)
  experiment at bnl},\ }\href
  {https://doi.org/https://doi.org/10.1016/j.nima.2007.06.023} {\bibfield
  {journal} {\bibinfo  {journal} {Nuclear Instruments and Methods in Physics
  Research Section A: Accelerators, Spectrometers, Detectors and Associated
  Equipment}\ }\textbf {\bibinfo {volume} {579}},\ \bibinfo {pages} {1096}
  (\bibinfo {year} {2007})}\BibitemShut {NoStop}%
\end{thebibliography}%

\end{document}